\documentclass[twocolumn,showpacs,preprintnumbers,amsmath,amssymb,superscriptaddress]{revtex4}

\usepackage{graphicx}
\usepackage{dcolumn}
\usepackage{bm,bbm}
\usepackage{color}
\usepackage{hyperref}
\usepackage[normalem]{ulem}

\font\scripti=cmmi7
\font\scriptscripti=cmmi5
\def\sib#1{\setbox0 = \hbox{\scripti #1}
  \kern-.02em\copy0\kern-\wd0
  \kern.04em\box0} % script italic bold 
\def\ssib#1{\setbox0 = \hbox{\scriptscripti #1}
  \kern-.02em\copy0\kern-\wd0
  \kern.04em\box0} % scriptscript italic bold
\font\tenib=cmmib10 % italic bold for math
\skewchar\tenib='177 \skewchar\tenib='177 \skewchar\tenib='177
\def\pbold#1{\setbox0 = \hbox{$ #1 $}
  \kern-.022em\copy0\kern-\wd0
  \kern.011em\copy0\kern-\wd0
  \kern.011em\copy0\kern-\wd0
  \kern.011em\copy0\kern-\wd0
  \kern.011em\box0} % poorman's bold

\renewcommand\d{\partial}

\newcommand\+{\dagger}
\newcommand\<{\langle}
\renewcommand\>{\rangle}

\renewcommand\Re{\mathrm{Re}}
\renewcommand\Im{\mathrm{Im}}

\newcommand\D{{\bm{D}}}

\renewcommand\H{\mathcal{H}}

\newcommand\J{{\bm{J}}}

\newcommand\p{{\bm{p}}}

\renewcommand\k{{\bm{k}}}

\newcommand\0{\mathbf{0}}
\renewcommand\r{{\bm{r}}}

\newcommand\R{{\bm{R}}}
\newcommand\X{{\bm{X}}}

\newcommand\up{\uparrow}
\newcommand\down{\downarrow}

\begin{document}
\title{Optical Spin Conductivity in Ultracold Quantum Gases}
\author{Yuta Sekino}
\affiliation{Quantum Hadron Physics Laboratory, RIKEN Nishina Center (RNC), Wako, Saitama, 351-0198, Japan}
\affiliation{Interdisciplinary Theoretical and Mathematical Sciences Program (iTHEMS), RIKEN, Wako, Saitama 351-0198, Japan}
\affiliation{RIKEN Cluster for Pioneering Research (CPR), Astrophysical Big Bang Laboratory (ABBL), Wako, Saitama, 351-0198 Japan}
\author{Hiroyuki Tajima}
\affiliation{Department of Mathematics and Physics, Kochi University, Kochi 780-8520, Japan}
\affiliation{Department of Physics, Graduate School of Science, The University of Tokyo, Tokyo 113-0033, Japan}
\author{Shun Uchino}
\affiliation{Advanced Science Research Center, Japan Atomic Energy Agency, Tokai, Ibaraki 319-1195, Japan}
\affiliation{Waseda Institute for Advanced Study, Waseda University, Shinjuku, Tokyo 169-8050, Japan}
\date{\today}

\begin{abstract}
We show that the optical spin conductivity being a small AC response of a bulk spin current and elusive in condensed matter systems can be measured in ultracold atoms.
We demonstrate that this conductivity contains rich information on quantum states by analyzing experimentally achievable systems such as a spin-1/2 superfluid Fermi gas, a spin-1 Bose-Einstein condensate, and a Tomonaga-Luttinger liquid.
The obtained conductivity spectra being absent in the Drude conductivity reflect quasiparticle excitations and non-Fermi liquid properties.
Accessible physical quantities include the superfluid gap and the contact for the superfluid Fermi gas, gapped and gapless spin excitations as well as quantum depletion for the Bose-Einstein condensate, and the spin part of the Tomonaga-Luttinger liquid parameter elusive in cold-atom experiments.
Unlike its mass transport counterpart, the spin conductivity serves as a probe applicable to clean atomic gases without disorder and lattice potentials.
Our formalism can be generalized to various systems such as spin-orbit coupled and nonequilibrium systems.
\end{abstract}
\maketitle

\section{\label{sec:introduction}Introduction}
Transport plays crucial roles in understanding states of matter in and out of equilibrium and paves the way to application such as control of matter and device fabrication.
In solid state physics, the main bearer of transport is an electron and the properties of the electric current have conventionally been investigated~\cite{mermin}.
Subsequently, the spin current being a flow of electric spin has attracted attention since the discovery of the giant magnetoresistance~\cite{binasch,baibich} and the tunneling magnetoresistance~\cite{miyazaki}.
More recently, due to the progress in nanofabrication technology of devices, physics in spin currents~\cite{maekawa} has also been widespread  over materials with spin-Hall effects~\cite{sinova}, and topological insulators~\cite{qi}.

One of the hot topics in the rapid growth of the spintronics is to measure AC spin currents in a direct manner~\cite{heinrich,woltersdorf,matsuo,jiao,sun,hahn,wei,weiler,li,kobayashi,kurimune}.
Such AC currents are detected in junction systems, and the determination of an AC conductivity of bulk spin transport is difficult in solid state systems.
To address this spin transport property, we shed light on ultracold atoms being an ideal platform for quantum simulation of many-body systems~\cite{schafer}.
Recently, the cold-atom analog of electronics referred to as atomtronics has attracted widespread attention~\cite{amico}, and transport measurements with ultracold atoms have been done with bulk~\cite{ott,strohmaier,Sommer:2011a,Sommer:2011b,Jepsen:2020,Bardon:2014,Hild:2014,Koschorreck:2013,schneider,ronzheimer,heinze,scherg,brown,nichols,Anderson:2019} and mesoscopic setups~\cite{krinner}.
One of the advantages of ultracold atoms is that spin-selective manipulation and probe are allowed, which opens up the possibility of precise measurements of spin transport~\cite{enss-thywissen}.

\begin{figure}[htbp]
\centering
\includegraphics[width=8cm]{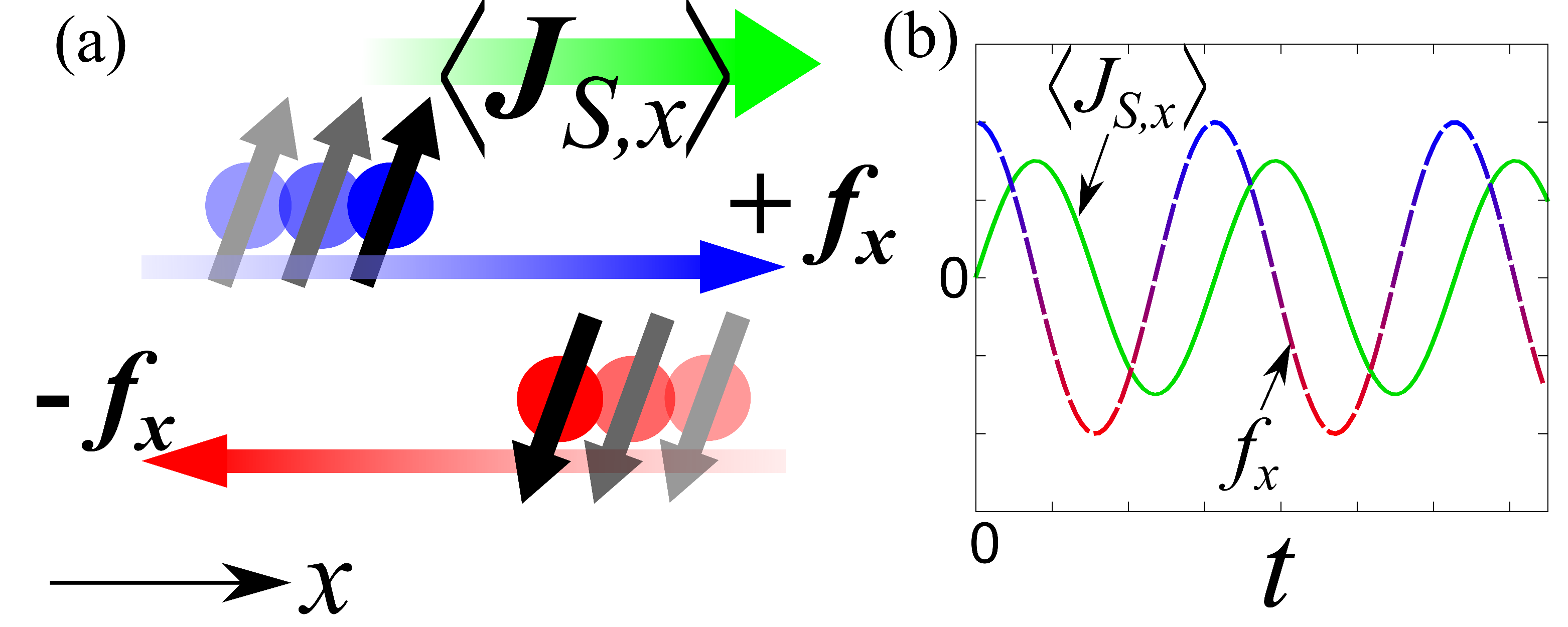}
\caption{\label{fig1}
(a) Schematic image for probing the optical spin conductivity $\sigma_{\alpha\beta}^{(S)}(\omega)$ in ultracold atomic gases with spin $S=1/2$.
The AC spin current $\<J_{S,x}(t)\>$ is induced by a spin-dependent driving forces $\pm f_x(t)$. (b) Time evolutions of $f_x(t)=F_x\cos(\omega_0t)$ and the induced spin current, which is related to $\sigma_{xx}^{(S)}(\omega=\omega_0)$ by $\<J_{S,x}(t)\>/F_x=\Re\,\sigma_{xx}^{(S)}(\omega_0)\cos(\omega_0t)+\Im\,\sigma_{xx}^{(S)}(\omega_0)\sin(\omega_0t)$ resulting from Eq.~(\ref{eq:X_S(t)}).}
\end{figure}
In this paper, we propose that ultracold atoms provide us with a simple way to measure an optical spin conductivity $\sigma_{\alpha\beta}^{(S)}(\omega)$, which characterizes an AC response of bulk spin transport.
The spectrum of $\sigma_{\alpha\beta}^{(S)}(\omega)$ includes richer information on bulk properties than its DC limit related to spin diffusion, which has been actively studied~\cite{Sommer:2011a,Sommer:2011b,Jepsen:2020,Bardon:2014,Hild:2014,Koschorreck:2013,enss-thywissen,Bulchandani:2021}.
This paper is organized as follows: In Sec.~\ref{sec:optical spin conductivity},
we provide the formalism of $\sigma_{\alpha\beta}^{(S)}(\omega)$ applicable to both continuum and optical lattice systems.
We then demonstrate the availability of $\sigma_{\alpha\beta}^{(S)}(\omega)$ by investigating experimentally verifiable systems.
Specifically, a spin-1/2 superfluid Fermi gas, a spin-1 Bose-Einstein condensate (BEC), and a Tomonaga-Luttinger (TL) liquid are studied in Secs.~\ref{sec:S=1/2}, \ref{sec:S=1}, and \ref{sec:TLL}, respectively.
Reflecting on nontrivial spin excitations, these systems show interesting transport properties being absent in the conventional Drude conductivity.
Section~\ref{sec:realization} is devoted to proposing a method to measure $\sigma_{\alpha\beta}^{(S)}(\omega)$ from an oscillating behavior in spin dynamics as shown in Fig.~\ref{fig1}.
Prospects towards spin-orbit coupled and nonequilibrium systems in Sec.~\ref{sec:extension} and other promising applications in Sec.~\ref{sec:applications} suggest that $\sigma^{(S)}_{\alpha\beta}(\omega)$ may be the Rosetta stone to unravel spin dynamics of various quantum many-body systems.
We conclude in Sec~\ref{sec:conclusion}.
In what follows, we set $\hbar=k_B=1$.

\section{\label{sec:optical spin conductivity}Optical spin conductivity}
We consider a spin-conserved system with spin $S=1/2,1,3/2,\cdots$, to which a time-dependent perturbation generating a pure spin current is applied.
Single-particle and interaction potentials can take arbitrary forms as long as spin is conserved (see Appendix~\ref{appendix:formalism} for details).
The time-dependent perturbation has the form of 
\begin{align}\label{eq:deltaH(t)}
\delta H_{\beta}(t)=-\int d\r\,f_\beta(t)r_\beta S_z(\r),
\end{align}
where $f_{\beta}(t)$ provides a driving force in the direction $\beta=x,y,z$ coupled to the spin density $S_z(\r)=\sum_{(s_z,i)}s_z\delta(\r-\r_{s_z,i})$ (see Fig.~\ref{fig1}).
Here, $\r_{s_z,i}$ is the position of the $i$th particle in the $s_z$ component with $s_z=-S,-S+1,\cdots,S$.
The spin current operator in the Heisenberg picture is given by $\bm{J}_S(t)=\sum_{(s_z,i)}s_z\frac{d\r_{s_z,i}(t)}{dt}$.
By the linear response theory, the optical spin conductivity $\sigma_{\alpha\beta}^{(S)}(\omega)$ with $\alpha=x,y,z$ is defined by
\begin{align}\label{eq:conductivity}
\<\tilde{J}_{S,\alpha}(\omega)\>=\sigma_{\alpha\beta}^{(S)}(\omega)\tilde{f}_{\beta}(\omega),
\end{align}
where $\tilde{J}_{S,\alpha}(\omega)$ and $\tilde{f}_\beta(\omega)$ are the Fourier transforms of $J_{S,\alpha}(t)$ and $f_\beta(t)$, respectively, and $\<\cdots\>$ denotes the expectation value with respect to the nonequilibrium state driven by $f_\beta(t)$~\cite{Note:Linear_response}.
We note that $\sigma_{\alpha\beta}^{(S)}(\omega)$ is the response not of a spin current density but of a total spin current.

We point out a difference from the mass current induced by a spin-independent perturbation~\cite{Wu:2015}.
In clean cold atomic gases trapped in a box or harmonic potential,  
the total center-of-mass motion is independent of quantum states of matter due to Kohn's theorem~\cite{Kohn:1961,Brey:1989,Li:1991}.
Thus, a system that breaks prior conditions of Kohn's theorem such as optical lattice or disordered systems must  be prepared to obtain a nontrivial mass response, which has recently been confirmed~\cite{Anderson:2019}.
In contrast,  the relative motion between spin components relevant to the optical spin conductivity can show a nontrivial response, once interatomic interactions are present.

We now provide two general properties of $\sigma_{\alpha\beta}^{(S)}(\omega)$ in a similar way as in the $S=1/2$ case~\cite{Enss:2012,Enss:2013,Note:S=1/2}.
First, the optical spin conductivity can be expressed in terms of a current-current correlation function $\chi_{\alpha\beta}(\omega)$:
\begin{align}\label{eq:conductivity1}
\sigma_{\alpha\beta}^{(S)}(\omega)&=\frac{i}{\omega^+}\left(\delta_{\alpha\beta}\sum_{s_z}\frac{s_z^2N_{s_z}}{m}+\chi_{\alpha\beta}(\omega)\right),
\end{align}
where $\omega^+=\omega+i0^+$, $m$ is a mass of a particle, $N_{s_z}$ is the particle number in the $s_z$ component, $\chi_{\alpha\beta}(\omega)=-i\int_{-\infty}^\infty dt\,e^{i\omega^+ t}\theta(t)\<[J_{S,\alpha}(t),J_{S,\beta}(0)]\>_0$ with the Heaviside step function $\theta(t)$, and $\<\cdots\>_0$ denotes the thermal average without the driving term.
Second, the frequency integral of the real part with $\alpha=\beta$ is exactly related to $N_{s_z}$ by the following $f$-sum rule~\cite{Note:Single-band}:
\begin{align}\label{eq:f-sum}
\int_{-\infty}^{\infty}\frac{d\omega}{\pi}\Re\,\sigma_{\alpha\alpha}^{(S)}(\omega)=\sum_{s_z}\frac{s_z^2 N_{s_z}}{m}.
\end{align}
Since this real part provides energy dissipation associated with spin excitations, we will from now on mainly focus on $\Re\,\sigma_{\alpha\alpha}^{(S)}(\omega)$~\cite{Note:Kramers-Kronig}.
To demonstrate what information can be captured by the spectrum of $\Re\,\sigma_{\alpha\alpha}^{(S)}(\omega)$, two homogeneous superfluids and a TL liquid at zero temperature are specifically addressed below.

\section{\label{sec:S=1/2}Spin-1/2 superfluid Fermi gas}
First, we investigate spin transport for a superfluid Fermi gas with $S=1/2$~\cite{Giorgini:2008,Zwerger:2012}.
By employing the mean-field theory~\cite{Eagles:1969,Leggett:1980}, $\Re\,\sigma_{\alpha\alpha}^{(S)}(\omega)$ is examined from a weakly interacting Bardeen-Cooper-Schrieffer (BCS) state to a Bose-Einstein condensate (BEC) of tightly-bound molecules.
In particular, we will show that the spin-singlet pairing results in the spectrum of $\Re\,\sigma_{\alpha\alpha}^{(S)}(\omega)$ quite different from that above the transition temperature, whose low-frequency behavior is well described by the conventional Drude conductivity~\cite{Enss:2012}.

In what follows, the $s_z=+1/2$ ($s_z=-1/2$) component is referred to as $\up$ ($\down$) and the spin-balanced case $N_\up=N_\down$ is considered.
The ground canonical Hamiltonian of this system is given by
\begin{align}\label{eq:H_fermion}
K&=\sum_\k\sum_{\sigma=\up,\down}(\varepsilon_\k-\mu)c_{\k,\sigma}^\+c_{\k,\sigma}\nonumber\\
&\quad-\frac{g}{\Omega}\sum_{\k,\p,\p'}c_{\k/2+\p,\up}^\+c_{\k/2-\p,\down}^\+c_{\k/2-\p',\down}c_{\k/2+\p',\up},
\end{align}
where $\varepsilon_\k=\k^2/(2m)$, $\mu$ is the chemical potential, $c_{\k,\sigma}$ is the annihilation operator of a Fermi atom with spin $\sigma$, and $\Omega$ is the volume of the system.
The coupling constant $g>0$ is related to the scattering length $a$ by $1/g=-m/(4\pi a)+\Omega^{-1}\sum_{|\k|<\Lambda}m/\k^2$ with the momentum cutoff $\Lambda$.

The optical spin conductivity within the mean-field theory can be analytically evaluated.
The spin current operator appearing in $\chi_{\alpha\beta}(\omega)$ is given by $\bm{J}_S=\sum_\k\frac{\k}{2m}(c_{\k,\up}^\+c_{\k,\up}-c_{\k,\down}^\+c_{\k,\down})$.
From the rotational symmetry of the system, $\sigma_{\alpha\alpha}^{(S)}(\omega)$ is independent of $\alpha=x,y,z$ and found to be (see Appendix~\ref{appendix:S=1/2} for details)
\begin{align}\label{eq:conductivity_fermions_1}
\Re\,\sigma_{xx}^{(S)}(\omega)=\sum_\k\frac{\pi\Delta^2k_x^2}{m^2|\omega|^3}\delta(|\omega|-2E_{\k,{\rm F}}),
\end{align}
where $\Delta>0$ is the superfluid order parameter and $E_{\k,{\rm F}}=\sqrt{(\varepsilon_\k-\mu)^2+\Delta^2}$ is the quasiparticle energy with momentum $\k$.
As the attraction becomes stronger, $\mu$ monotonically decreases from $\mu\simeq E_\mathrm{F}=k_\mathrm{F}^2/(2m)$ in the BCS limit [$(k_{\rm F}a)^{-1}\to-\infty$] to $\mu\simeq-1/(2ma^2)$ in the BEC limit [$(k_{\rm F}a)^{-1}\to+\infty$], where a dimensionless parameter $(k_{\rm F}a)^{-1}$ given by a Fermi momentum $k_{\mathrm{F}}$  and the scattering length $a$ characterizes the strength of the attraction~\cite{Note:mu}.
Performing the integration over $\k$ in Eq.~(\ref{eq:conductivity_fermions_1}), we obtain
\begin{align}
\label{eq:conductivity_fermions_2}
\Re\,\sigma_{xx}^{(S)}(\omega)
&=\frac{\sqrt{m}\Delta^2\Omega}{12\pi}\frac{[\varepsilon_+(\omega)]^{\frac32}+\theta(\varepsilon_-(\omega))[\varepsilon_-(\omega)]^{\frac32}}{\omega^2\sqrt{\omega^2-4\Delta^2}}\nonumber\\
&\quad\times\theta(|\omega|-2E_\mathrm{gap}),
\end{align}
where $E_\mathrm{gap}\equiv\min_\k(E_{\k,{\rm F}})=\Delta\,\theta(\mu)+\sqrt{\mu^2+\Delta^2}\,\theta(-\mu)$ is the energy gap and $\varepsilon_\pm(\omega)\equiv2\mu\pm \sqrt{\omega^2-4\Delta^2}$.
Note that $\varepsilon_-(\omega)$ is relevant for $2\Delta<|\omega|<2\sqrt{\mu^2+\Delta^2}$ with $\mu>0$.

\begin{figure}[t]
\centering
\includegraphics[width=7cm]{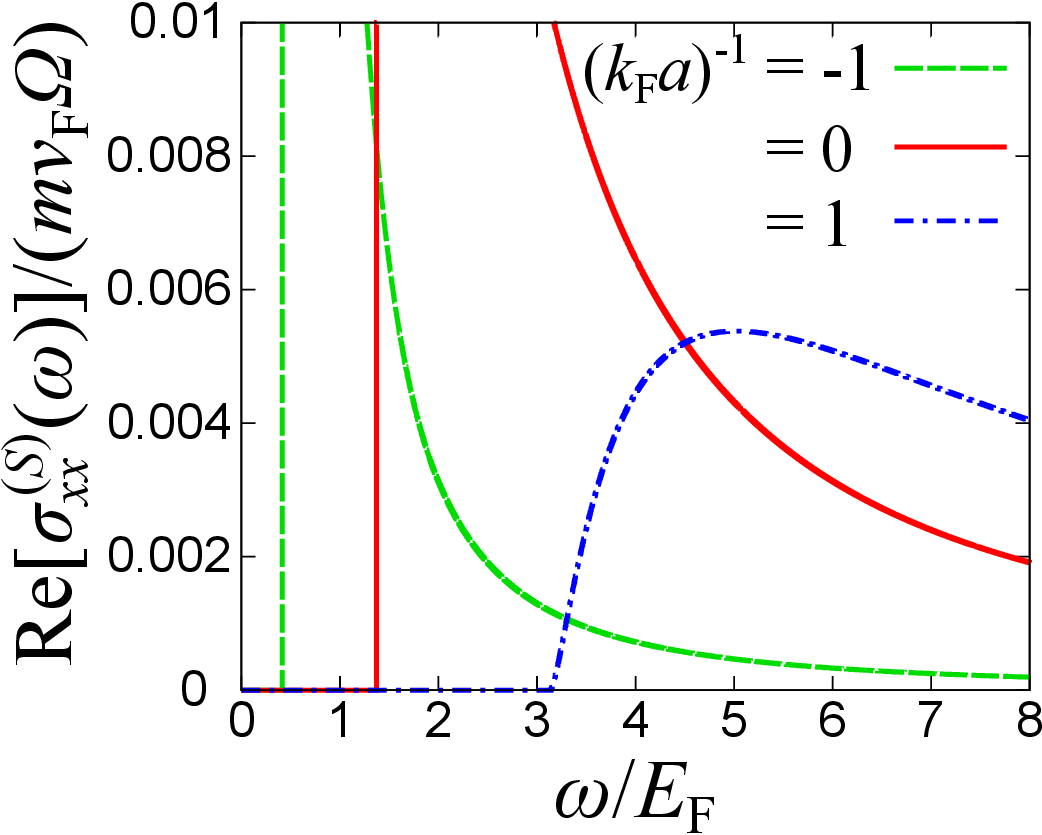}
\caption{\label{fig:2}
Spectra of the optical spin conductivity in a spin-1/2 superfluid Fermi gas at zero temperature.
The conductivities vanish for small $\omega$ due to the spin-singlet pairing.
At $(k_{\rm F}a)^{-1}=-1,0$, positive chemical potentials result in coherence peaks near the thresholds.
Here, $v_{\rm F}=k_{\rm F}/m$ is the Fermi velocity and $\Omega$ is the volume.}
\end{figure}
Equation~(\ref{eq:conductivity_fermions_1}) clarifies that the structure of the quasiparticle spectrum strongly affects $\Re\,\sigma_{xx}^{(S)}(\omega)$.
It is notable that $\Re\,\sigma_{xx}^{(S)}(\omega)$ vanishes for $|\omega|<2E_\mathrm{gap}$ as shown by $\theta(|\omega|-2E_\mathrm{gap})$ in Eq.~\eqref{eq:conductivity_fermions_2}.
This reflects the fact that spin excitations are associated with the dissociation of spin-singlet Cooper pairs or molecules and require energy being larger than $2E_\mathrm{gap}$.
Figure~\ref{fig:2} shows the spectra of the optical spin conductivity for different interaction strengths.
The behavior of $\Re\,\sigma_{xx}^{(S)}(\omega)$ near the threshold ($|\omega|\to2E_\mathrm{gap}+0$) depends on the sign of the chemical potential~\cite{Note:mu}:
\begin{align}
\Re\,\sigma_{xx}^{(S)}(\omega)=
\begin{cases}
\frac{\Omega}{12\pi}\sqrt{\frac{m\mu^3}{2\Delta\delta\omega}}&(\mu>0)\\
\frac{\Delta^2\Omega}{48\pi}\sqrt{\frac{m\delta\omega^3}{\sqrt{\mu^2+\Delta^2}|\mu|^5}}&(\mu<0)
\end{cases}
\end{align}
with $\delta\omega=|\omega|-2E_\mathrm{gap}\to+0$.
In the case of $\mu>0$ [$(k_\mathrm{F}a)^{-1}=-1,0$ in Fig.~\ref{fig:2}], the flat band at $|\k|=\sqrt{2m\mu}$ results in the divergent behavior $\Re\,\sigma_{xx}^{(S)}(\omega)\sim1/\sqrt{\delta\omega}$, which is the so-called coherence peak~\cite{Schrieffer:1964}.
On the other hand, $E_{\k,{\rm F}}$ is a monotonically increasing function of $|\k|$ on the BEC side with $\mu<0$ [$(k_\mathrm{F}a)^{-1}=1$ in Fig.~\ref{fig:2}] and the optical spin conductivity decreases as $\Re\,\sigma_{xx}^{(S)}(\omega)\sim(\delta\omega)^{3/2}$.
In this way, the optical spin conductivity proves the excitation properties and the aspects of a spin insulator in the superfluid Fermi gas.
We emphasize that these transport properties of the superfluid cannot be captured by the conventional Drude conductivity, whose real part takes a Lorentz distribution.

We mention the validity of the mean-field analysis.
For any $(k_\mathrm{F}a)^{-1}$, our result in Eq.~(\ref{eq:conductivity_fermions_1}) satisfies exact relations such as the $f$-sum rule in Eq.~(\ref{eq:f-sum}) and the high-frequency tail $\Re\,\sigma_{xx}^{(S)}(\omega)=C\Omega/[12\pi (m|\omega|)^{3/2}]$~\cite{Enss:2012,Note:S=1/2,Hofmann:2011} with Tan's contact $C$~\cite{Tan:2008}.
Indeed, the high-frequency asymptotics of Eq.~\ref{eq:conductivity_fermions_2} provides the mean-field value of the contact $C=m^2\Delta^2$.
In addition, the mean-field theory employed in this paper gives semi-quantitative descriptions of physical quantities throughout the BCS-BEC crossover at zero temperature regardless of the presence of the strong interaction~\cite{Horikoshi,Note:mean-field}.

\section{\label{sec:S=1}Spin-1 polar condensate}
The optical spin conductivity can also be useful for bosonic systems.
To see this, we next investigate spin transport for a spin-1 BEC within the Bogoliubov theory~\cite{kawaguchi,Note:Bogoliubov}.
Focusing on the polar phase realized with $^{23}$Na and $^{87}$Rb~\cite{Stamper-Kurn:2013}, we will show a nontrivial AC spin response, which is again different from the Drude conductivity.
In this phase, bosons condense only in the $s_z=0$ channel, which is decoupled from the spin channels ($s_z=\pm1$)~\cite{Uchino:2010}.
By definition of $\bm{J}_S(t)$, only quasiparticles in the spin channels contribute to spin transport and the $f$ sum of $\Re\,\sigma_{xx}^{(S)}(\omega)$ in Eq.~(\ref{eq:f-sum}) is related to the particle number $N_{1}+N_{-1}$ in the spin channels arising from quantum depletion~\cite{Uchino:2010}.

The grand canonical Hamiltonian of the system is given by~\cite{Uchino:2010}
\begin{align}\label{eq:H_boson}
K&=\sum_\k\sum_{s_z=0,\pm1}(\varepsilon_\k+qs_z^2-\mu)a_{\k,s_z}^\+a_{\k,s_z}\cr
&\quad+\frac{c_0}{2\Omega}\sum_{\k,\p,\p'}\sum_{s_z,s_z'}a_{\k/2+\p,s_z}^\+a_{\k/2-\p,s_z'}^\+a_{\k/2-\p',s_z}a_{\k/2+\p',s_z'}\cr
&\quad+\frac{c_1}{2\Omega}\sum_{\k,\p,\p'}\sum_{s_z,s_z',s_z'',s_z'''}\mathbf{S}_{s_z,s_z'}\cdot\mathbf{S}_{s_z'',s_z'''}\cr
&\quad\times\,a_{\k/2+\p,s_z}^\+a_{\k/2-\p,s_z''}^\+a_{\k/2-\p',s_z'}a_{\k/2+\p',s_z'''},
\end{align}
where $q$ characterizes the quadratic Zeeman effect~\cite{Stamper-Kurn:2013}, $\mu$ is the chemical potential, $a_{\k,s_z}$ is the annihilation operator of a Bose atom with spin $s_z$, and $\Omega$ is the volume of the system.
In a spin-1 BEC, the interatomic interactions can be characterized by the spin-independent coupling constant $c_0>0$ and spin-dependent coupling constant $c_1$~\cite{Ho:1998,Ohmi:1998}.
The spin-1 matrices $\mathbf{S}_{s_z,s_z'}=(\mathrm{S}^x_{s_z,s_z'},\mathrm{S}^y_{s_z,s_z'},\mathrm{S}^z_{s_z,s_z'})$ are given by
\begin{align}
\mathrm{S}^x&=\frac{1}{\sqrt2}
\begin{pmatrix}
0&1&0\\
1&0&1\\
0&1&0
\end{pmatrix},\
\mathrm{S}^y=\frac{i}{\sqrt2}
\begin{pmatrix}
0&-1&0\\
1&0&-1\\
0&1&0
\end{pmatrix},\\
\mathrm{S}^z&=
\begin{pmatrix}
1&0&0\\
0&0&0\\
0&0&-1
\end{pmatrix}.
\end{align}

The optical spin conductivity within the Bogoliubov theory can be analytically evaluated.
The spin current operator appearing in $\chi_{\alpha\beta}(\omega)$ is given by $\bm{J}_S=\sum_\k\frac{\k}{m}(a_{\k,1}^\+a_{\k,1}-a_{\k,-1}^\+a_{\k,-1})$.
From the rotational symmetry of the system, $\sigma_{\alpha\alpha}^{(S)}(\omega)$ is independent of $\alpha=x,y,z$ and found to be (see Appendix~\ref{appendix:S=1} for details)
\begin{align}\label{eq:conductivity_bosons_1}
\Re\,\sigma_{xx}^{(S)}(\omega)=\sum_\k\frac{4\pi n_0^2c_1^2k_x^2}{m^2|\omega|^3}\delta(|\omega|-2E_{\bm{k},s}),
\end{align}
where $E_{\bm{k},s}=\sqrt{(\varepsilon_\k+q)(\varepsilon_\k+q+2n_0c_1)}$ is the quasiparticle energy in the spin channels and $n_0$ is the condensate fraction.
In the polar phase satisfying $q+n_0c_1>n_0|c_1|$, the spin excitations are gapped with the spin gap $E_\mathrm{gap}=\sqrt{q(q+2n_0c_1)}$, while the gap is closed on boundaries of the phase ($q+n_0c_1=n_0|c_1|$).
From Eq.~(\ref{eq:conductivity_bosons_1}), the optical spin conductivity is sensitive to whether spin excitations are gapped or gapless.
By performing the integration over $\k$ in Eq.~(\ref{eq:conductivity_bosons_1}), the analytical form of $\Re\,\sigma_{xx}^{(S)}(\omega)$ is found to be
\begin{align}\label{eq:conductivity_bosons_2}
\Re\,\sigma_{xx}^{(S)}(\omega)=\frac{\sqrt{m}n_0^2c_1^2\Omega}{3\pi}\frac{[\varepsilon_s(\omega)]^{\frac32}}{\omega^2\sqrt{\omega^2+4n_0^2c_1^2}}\theta(|\omega|-2E_\mathrm{gap})
\end{align}
with $\varepsilon_s(\omega)\equiv\sqrt{\omega^2+4n_0^2c_1^2}-2(q+n_0c_1)$.

\begin{figure}[t]
\centering
\includegraphics[width=7cm]{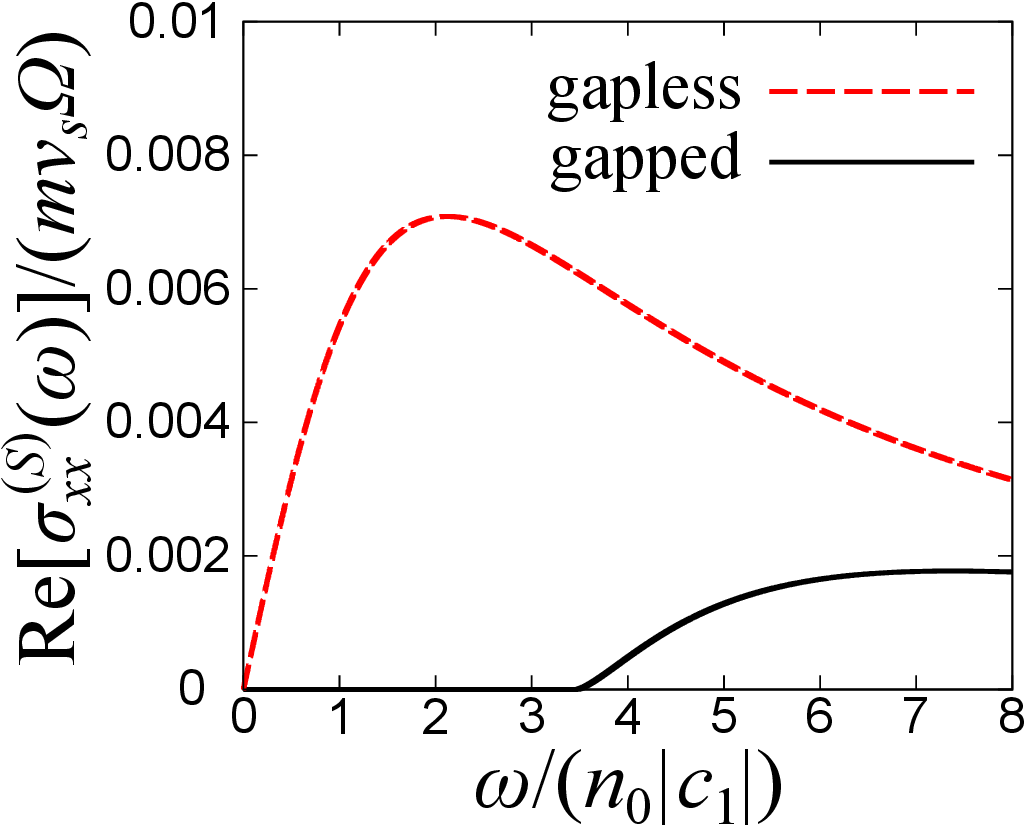}
\caption{\label{fig:3}
Optical spin conductivity spectra in a spin-1 BEC at zero temperature.
The solid line shows the result in the spin-gapped case ($q=n_0c_1$), where the conductivity becomes nonzero for $\omega>2E_\mathrm{gap}$, while the dashed line shows that in the gapless case ($q+n_0c_1=n_0|c_1|$).
Note that $v_s=\sqrt{n_0|c_1|/m}$ is associated with spin excitations and different from the sound velocity.}
\end{figure}
Figure~\ref{fig:3} shows $\Re\,\sigma_{xx}^{(S)}(\omega)$ of the polar BEC.
Inside the polar phase with the spin gap, $\Re\,\sigma_{xx}^{(S)}(\omega)$ vanishes for $|\omega|<2E_\mathrm{gap}$ and its decreasing behavior near the threshold ($|\omega|\to2E_\mathrm{gap}+0$) is similar to that of the superfluid Fermi gas on the BEC side because of the similarity between $E_{\bm{k},s}$ and $E_{\k,{\rm F}}$.
On the other hand, $\Re\,\sigma_{xx}^{(S)}(\omega)$ on the phase boundaries with gapless spin excitations behaves linearly $\Re\,\sigma_{xx}^{(S)}(\omega)/\Omega=|\omega|/(48\pi v_s)$ in the low-frequency region ($|\omega|\ll n_0|c_1|$) with the spin velocity $v_s=\sqrt{n_0|c_1|/m}$ (see Fig.~\ref{fig:3}).
We note that this linear behavior is quite different from that of the Drude conductivity for conventional spin-gapless systems.
In addition, by taking $\omega\to\pm\infty$ in Eq.\eqref{eq:conductivity_bosons_2}, one can find the power-law tail $\Re\,\sigma_{xx}^{(S)}(\omega)=\sqrt{m}n_0^2c_1^2\Omega/(3\pi|\omega|^{3/2})$ in a similar way as for the superfluid Fermi gas.
In this way, the obtained optical spin conductivity reflects the properties of spin excitations inherent to the polar BEC.

\section{\label{sec:TLL}Tomonaga-Luttinger liquids}
In the case of the charge response, the optical conductivity is known to be a powerful tool for characterization of non-Fermi liquids where there is no well-defined quasiparticle~\cite{sachdev,chowdhury}.
To demonstrate the usefulness of the spin counterpart, here we consider one-dimensional quantum critical states with spin-1/2 where TL liquids being typical non-Fermi liquids are realized.
In such states, the low-energy properties can be explained by an effective Hamiltonian where charge and spin degrees of freedom are separated, and therefore information of the spin part is expected to be captured by spin transport.

The effective Hamiltonian $H=H_{C}+H_S$ of TL liquids is bosonized as~\cite{giamarchi}
\begin{subequations}\label{eq:H_C/S}
\begin{align}
H_C&=\frac{1}{2\pi}\int dx\,u_C[K_C(\pi \Pi_C(x))^2+(\nabla\phi_C(x))^2/K_C],\\
H_S&=\frac{1}{2\pi}\int dx\,u_S[K_S(\pi \Pi_S(x))^2+(\nabla\phi_S(x))^2/K_S]\nonumber\\
&\quad+\frac{2g}{(2\pi\alpha)^2}\int dx\,\cos(\sqrt{8}\phi_S(x)).
\end{align}
\end{subequations}
Here, $H_{C(S)}$ is  the Hamiltonian of the charge (spin) degree of freedom that consists of
bosonic fields $\Pi_{C(S)}$ and $\phi_{C(S)}$ satisfying the canonical commutation relation:
\begin{eqnarray}
[\phi_{C(S)}(x_1),\Pi_{C(S)}(x_2)]=i\delta(x_1-x_2).
\end{eqnarray}
Physically, $\phi_{C(S)}$  and $\Pi_{C(S)}$  describe the charge (spin) density fluctuations and their conjugate phase fluctuations, respectively.
In addition, $u_{C(S)}$ and $K_{C(S)}$ represent the velocity and the TL parameter in the charge (spin) component, respectively,
$g$ is the coupling constant arising from the back-scattering process, and $\alpha$ is the cutoff of the low-energy theory~\cite{Note:TLL}.
We note that the above Hamiltonian covers low-energy behaviors of experimentally important systems such as the Fermi-Hubbard model and Yang-Gaudin model.

To extract the essential feature, we perform bosonization allowing to express the spin current operator as $\sqrt{2}u_{S}K_{S}\Pi_{S}(x)$. 
In addition, $\cos(\sqrt{8}\phi_S(x))$ in $H_{C}$ does not commute with the spin current operator and generates nontrivial conductivity spectra. The basic feature of the spin conductivity is calculated with the memory function method~\cite{giamarchi1991,giamarchi1992} (see Appendix~\ref{appendix:TL} for details).
At zero temperature, this leads to
\begin{align}\label{eq:conductivity_TLL}
\Re\,\sigma^{(S)}(\omega)\propto\omega^{4K_S-5}.
\end{align}
Thus, we find that the optical spin response is powerful in that $K_{S}$ essential to the critical properties yet elusive in cold atoms is determined by the frequency dependence.
We also note that the similar power law dependence in the spin conductivity shows up for a spin insulating system where the back scattering process is relevant.
In this case, the spin conductivity spectrum vanishes at frequencies below the spin gap yet obeys the power-law behavior at frequencies above the gap~\cite{giamarchi}.

\section{\label{sec:realization}Experimental realization}
We now discuss how to measure $\sigma_{\alpha\beta}^{(S)}(\omega)$ in experiments.
For ultracold atomic gases, we have several ways to induce the perturbation in Eq.~(\ref{eq:deltaH(t)}).
The most straightforward way is to apply a time-dependent gradient of a magnetic field $B(\r,t)\propto f_\beta(t)r_\beta$ along the $z$ axis~\cite{medley,Jotzu:2015}.
Such a gradient potential can also be produced by the optical Stern-Gerlach effect~\cite{Taie:2010}.
Furthermore, ultracold atoms allow us to directly observe $\<\bm{J}_S(t)\>=\frac{d\<\X_S(t)\>}{dt}$ because
\begin{align}
\<\X_S(t)\>\equiv\int d\r\,\r\<S_z(\r,t)\>
\end{align}
is given in terms of the observable spin density $\<S_z(\r,t)\>$~\cite{medley,valtolina}.
Hereafter, we focus on $\<\X_S(t)\>$ rather than $\<{\bm J}_S(t)\>$ to measure $\sigma_{\alpha\beta}^{(S)}(\omega)$.

In order to provide a concrete scheme of measurement, we consider a case with a single-frequency driving $f_\beta(t)=F_\beta\cos(\omega_0t)$.
In this case, $\<X_{S,\alpha}(t)\>$ shows an oscillating behavior, which is exactly related to $\sigma_{\alpha\beta}^{(S)}(\omega=\omega_0)$ by
\begin{align}\label{eq:X_S(t)}
&\frac{\<\delta X_{S,\alpha}(t)\>}{F_\beta}\nonumber\\
&=-\frac{\Im\,\sigma_{\alpha\beta}^{(S)}(\omega_0)}{\omega_0}\cos(\omega_0t)+\frac{\Re\,\sigma_{\alpha\beta}^{(S)}(\omega_0)}{\omega_0}\sin(\omega_0t)
\end{align}
with $\<\delta X_{S,\alpha}(t)\>\equiv\<X_{S,\alpha}(t)\>-\<X_{S,\alpha}\>_0$ (see Appendix~\ref{sec:single_frequency} for details).
Thus, $\sigma_{\alpha\beta}^{(S)}(\omega_0)$ can be extracted through the oscillation analysis of $\<X_{S,\alpha}(t)\>$~\cite{Note:Measurement_proposals}.
We next discuss feasibility in realistic experiments.
An accessible region of $\omega_0$ depends on the way to generate $f_\beta(t)$.
For a magnetic-field gradient, it is feasible up to frequencies of the order of kHz while noises coming from back electromotive forces to coils and metallic chambers would become significant in higher $\omega_0$~\cite{Nakajima:private_communications}.
On the other hand, a higher frequency region can easily be accessed by an optical driving force.
There, the lower bound of $\omega_0$ would be determined so as to avoid heating coming from photon scatterings~\cite{Nakajima:private_communications}.
In the case of the spin-1/2 Fermi superfluid, a typical many-body scales such as $\Delta$ is about $1$kHz, which can be accessible with both magnetic and optical gradients (see e.g. Ref.~\cite{Biss:2022} and references therein). 
In addition, noises to the magnitude of $\sigma_{\alpha\beta}^{(S)}(\omega)$ would arise from measurement of $\<S_z(\r,t)\>$ and the accuracy of $F_\alpha$ in each cycle.

\section{\label{sec:extension}Extensions of the results and proposed method}
We now discuss generalizations of our results [Eqs.~(\ref{eq:conductivity1}) and (\ref{eq:f-sum})] and proposal to measure $\sigma_{\alpha\beta}^{(S)}(\omega)$ based on Eq.~(\ref{eq:X_S(t)}).
First, we point out that Eqs.~(\ref{eq:conductivity1}) and (\ref{eq:f-sum}) as well as the measurement scheme for $S=1/2$ is naturally extended to two-component bosons~\cite{Myatt:1996,Hall:1998a,Hall:1998b} by regarding two species as spin-up and spin-down states.

We next discuss an extension of the driving force to measure $\sigma_{\alpha\beta}^{(S)}(\omega)$.
While generation of a pure spin current was considered so far, a more general time-dependent potential gradient inducing both mass and spin currents may be more practical in some experimental setups.
For example, such a perturbation has been realized by applying a magnetic-field gradient to $^{40}$K atoms with two hyperfine states~\cite{Jotzu:2015}. 
We find that this kind of perturbation is also available to measure $\sigma_{\alpha\beta}^{(S)}(\omega)$ when investigating spin-1/2 gases confined by harmonic or box traps (see Appendix~\ref{appendix:spin_mass} for details).
In particular, in the spin-balanced case, cross correlations between mass and spin vanish, so that Eq.~(\ref{eq:X_S(t)}) holds even when both mass and spin currents are induced.
The extension of our measurement scheme to spin-imbalanced cases is also possible.

Our results can also be generalized to spin-orbit coupled systems without spin conservation.
Spin transport in these systems has been actively studied not only in spintronics~\cite{sinova} but also in cold-atom experiments~\cite{galitski,Li:2019,spielman}.
Even in such systems, $\sigma_{\alpha\beta}^{(S)}(\omega)$ satisfying Eqs.~(\ref{eq:conductivity1}) and (\ref{eq:f-sum}) can be defined and measured by simply generalizing our proposed method.
To see this, we consider two-component gases with spin-orbit and Rabi couplings realized with ultracold atoms~\cite{Lin:2011,Wang:2012,Cheuk:2012} (see Appendix~\ref{appendix:SOC} for details).
For convenience, we define $X_{S,\alpha}^{\hat{a}}\equiv\int d\r\,r_\alpha S_{\hat{a}}(\r)$ and $\delta H_\beta^{\hat{b}}(t)\equiv-f_\beta^{\hat{b}}(t)X_{S,\beta}^{\hat{b}}$, where $S_{\hat{a}}(\r)$ is the spin density operator along the direction $\hat{a}=x,y,z$ in spin space.
We emphasize that $\<X_{S,\alpha}^{\hat{a}}(t)\>$ can be measured by observing the spin density.
In the spin-conserved case, Eq.~(\ref{eq:X_S(t)}) relates $\sigma_{\alpha\beta}^{(S)}(\omega)$ to the oscillating $\<X_{S,\alpha}^z(t)\>$ under the perturbation $\delta H_\beta^z(t)$ [Eq.~(\ref{eq:deltaH(t)})].
In the presence of the spin-orbit and Rabi couplings, $\sigma_{\alpha\beta}^{(S)}(\omega)$ is rewritten in terms of the thermal average of $S_{x}(\r)$ and four responses of $\<X_{S,\alpha}^{\hat{a}}(t)\>$ under $\delta H_\beta^{\hat{b}}(t)$ with $\hat{a},\hat{b}\in\{y,z\}$ [see Eq.~\eqref{eq_app:conductivity_SOC_result}].
These responses can be measured from oscillation analyses similar to Eq.~(\ref{eq:X_S(t)}).

Finally, as in the charge response~\cite{Shimizu:2010}, Eqs.~(\ref{eq:conductivity1}) and (\ref{eq:f-sum}) can be generalized to nonequilibrium systems with spin conservation, whose optical spin conductivity is measurable from spin dynamics in cold-atom experiments
(see Appendix~\ref{appendix:nonequilibrium} for details).
Unlike in equilibrium, the nonequilibrium spin conductivity can have a negative real part associated with energy gain~\cite{Tsuji:2009}.

\section{\label{sec:applications}Other promising applications}
The optical spin conductivity has a variety of prospects.
The potential of $\sigma_{\alpha\beta}^{(S)}(\omega)$ to detect a topological phase transition has been recently pointed out~\cite{Tajima:2021}.
Our formalism and proposal are applicable to optical lattice systems. 
For instance, it is important to confirm an anomalous frequency dependence in $\sigma^{(S)}(\omega)$ of spin chains~\cite{Agrawal:2020} whose spin superdiffusion has attracted attention~\cite{Bulchandani:2021,Wei:2021}.
As the optical charge conductivity measurement has already served as valuable probes for pseudogap phenomena~\cite{Homes:1993}, non-Fermi liquids~\cite{chowdhury,sachdev}, and photoinduced insulator-metal transitions~\cite{Iwai:2003,Cavalleri:2004,Okamoto:2007}, $\sigma_{\alpha\beta}^{(S)}(\omega)$ is expected to be a key quantity to understand spin dynamics of strongly correlated and nonequilibrium systems.
For instance, unlike photoemission spectroscopy conventionally used to study pseudogaps of ultracold atoms~\cite{Stewart:2008,Gaebler:2010,Sagi:2015}, $\sigma_{\alpha\beta}^{(S)}(\omega)$ involves both upper and lower branches of single-particle excitations, so that measurement of $\sigma_{\alpha\beta}^{(S)}(\omega)$ would deepen our understanding of the pseudogap phenomenon~\cite{Mueller:2017}.
In terms of unconventional quantum liquids, it is also interesting to investigate a spin liquid~\cite{Savary:2016}, which has recently been realized with cold atoms~\cite{Semeghini:2021}.
Finally, applications to nonequilibrium states such as Floquet time crystals~\cite{Else:2016} and nonlinear responses such as shift spin currents~\cite{morimoto} are also 
promising routes for the spin conductivity.

\section{\label{sec:conclusion}Conclusion}
In this paper, we discussed the optical spin conductivity $\sigma_{\alpha\beta}^{(S)}(\omega)$, which serves as a valuable probe to examine many-body interacting systems with spin degrees of freedom and can be measured with existing methods in cold-atom experiments.
First, the formalism of $\sigma_{\alpha\beta}^{(S)}(\omega)$ applicable to both continuum and optical lattice systems was provided.
We then theoretically investigated three systems to show the availability of the optical spin conductivity.
For the superfluid Fermi gas, the gapped single-particle excitations result in the gap of the spectrum $\Re\,\sigma_{xx}^{(S)}(\omega)$ and the flat band for $\mu>0$ leads to the coherence peak.
For the spinor BEC, $\Re\,\sigma_{xx}^{(S)}(\omega)$ detects gapped spin excitations in the polar phase as well as gapless spin excitations on the phase boundaries, and its $f$ sum is related to quantum depletion.
In addition, from Eq.~\eqref{eq:conductivity_TLL}, the optical spin conductivity is found to be related to the spin part of the TL liquid parameter $K_S$ elusive in cold-atom experiments.
We also proposed that the optical spin conductivity can be measured from the oscillation in spin dynamics [Eq.~\eqref{eq:X_S(t)}].
This proposal can be extended to various cases including spin-orbit coupled and nonequilibrium systems.
As mentioned in Sec.~\ref{sec:applications}, various applications of the optical spin conductivity as probes for exotic spin dynamics are promising.

\noindent{\it Note added}---
When this paper was being finalized, there appeared a paper~\cite{Carlini:2021}, where a spin drag effect and related $f$-sum rules due to a spin-dependent perturbation are discussed.

\acknowledgments
The authors thank H.~Konishi, M.~Matsuo, S.~Nakajima, and Y.~Nishida for useful discussions.
YS is supported by JSPS KAKENHI Grants No.~19J01006 and Pioneering Program of RIKEN for Evolution of Matter in the Universe (r-EMU).
HT is supported by Grant-in-Aid for Scientific Research provided by JSPS through No.~18H05406.
SU is supported by MEXT Leading Initiative for Excellent Young Researchers and Matsuo Foundation.

\appendix
\section{\label{appendix:formalism}Formalism with spin conservation}
We consider spin transport in a system with spin $S=1/2,1,3/2,\cdots$.
Here, we employ the first quantization formalism to clarify the connection between the spin current and the spin-resolved center-of-mass motion.
The Hamiltonian of the system is given by $\H(t)=H+\delta H_{\beta}(t)$.
The nonperturbative term $H$ is given by
\begin{align}\label{eq_app:H}
H=\sum_{(s_z,i)}\left(\frac{1}{2m}\left[\p_{s_z,i}-{\bm{A}}_{s_z}(\r_{s_z,i})\right]^2+V_{s_z}(\r_{s_z,i})\right)+H_{\rm int},
\end{align}
where $m$ is a mass of a particle and labels of particles take $s_z=-S,-S+1,\cdots,S$ and $i=1,2,\cdots,N_{s_z}$ with $N_{s_z}$ being the particle number in the $s_z$ component.
The operators $\r_{s_z,i}$ and $\p_{s_z,i}$ denote the coordinate and momentum operators of the particle with a label $(s_z,i)$ and they satisfy the following canonical commutation relations:
\begin{subequations}\label{eq_app:[r,p]}
\begin{align}
[(r_{s_z,i})_\alpha,(r_{s_z',i'})_\beta]&=[(p_{s_z,i})_\alpha,(p_{s_z',i'})_\beta]=0,\\
[(r_{s_z,i})_\alpha,(p_{s_z',i'})_\beta]&=i\delta_{s_zs_z'}\delta_{ii'}\delta_{\alpha\beta},
\end{align}
\end{subequations}
where $\alpha$ and $\beta$ denote Cartesian components $x,y,z$ in the coordinate space.
The functions ${\bm{A}}_{s_z}(\r)$ and $V_{s_z}(\r)$ are spin-dependent vector and scalar potentials, respectively, and the interaction term $H_{\rm int}$ is assumed to be described by pairwise potentials and commute with the $z$ component of the spin density operator $S_z(\r)=\sum_{(s_z,i)}s_z\delta(\r-\r_{s_z,i})$.
The time-dependent perturbation term is given by
\begin{align}\label{eq_app:deltaH(t)}
\delta H_\beta(t)=-\int d\r\,f_\beta(t)r_\beta S_z(\r)=-f_\beta(t)X_{S,\beta},
\end{align}
where $f_\beta(t)$ provides a driving force in the direction $\beta$ coupled to the spin density $S_z(\r)$.
The operator $\X_S\equiv\int d\r\,\r S_z(\r)$ measures the coordinate characterizing spin dynamics driven by $f_\beta(t)$.
In terms of the spin-resolved center-of-mass coordinate $\bm{R}_{s_z}=\sum_{i=1}^{N_{s_z}}\r_{s_z,i}/N_{s_z}$, $\X_S$ is rewritten as $\X_S=\sum_{s_z}s_zN_{s_z}\R_{s_z}$.
The perturbation generates a spin current, whose corresponding operator in the Heisenberg picture is given by $\bm{J}_S(t)=\sum_{(s_z,i)}s_z\frac{d\r_{s_z,i}(t)}{dt}$.
This operator can be rewritten as
\begin{align}\label{eq_app:J_S(t)}
\bm{J}_S(t)=\frac{d\X_{S}(t)}{dt}=\sum_{(s_z,i)}\frac{s_z}{m}[\p_{s_z,i}(t)-{\bm{A}}_{s_z}(\r_{s_z,i}(t))].
\end{align}

\subsection{Optical spin conductivity}
We now derive the expression of the optical spin conductivity $\sigma_{\alpha\beta}^{(S)}(\omega)$ [Eq.~\eqref{eq:conductivity1}] in terms of a retarded response function for a spin current.
The optical spin conductivity $\sigma_{\alpha\beta}^{(S)}(\omega)$ is given as the linear response of the spin current to the driving force:
\begin{align}\label{eq_app:conductivity}
\<\tilde{J}_{S,\alpha}(\omega)\>=\sigma_{\alpha\beta}^{(S)}(\omega)\tilde{f}_{\beta}(\omega),
\end{align}
where $\tilde{J}_{S,\alpha}(\omega)$ and $\tilde{f}_\alpha(\omega)$ are the Fourier transforms of $J_{S,\alpha}(t)$ and $f_\alpha(t)$, respectively, and $\<\cdots\>$ denotes the expectation value with respect to a nonequilibrium state driven by the external force.
The Kubo formula provides 
\begin{align}\label{eq_app:conductivity0}
\sigma_{\alpha\beta}^{(S)}(\omega)=i\int_{-\infty}^\infty\!\!dt\,e^{i\omega^+ t}\theta(t)\<[J_{S,\alpha}(t),X_{S,\beta}(0)]\>_0,
\end{align}
where $\omega^+\equiv\omega+i0^+$, $\theta(t)$ is the Heaviside step function, and $\<\cdots\>_0$ denotes the thermal average without the external force.
From Eq.~(\ref{eq_app:J_S(t)}) and $\<[J_{S,\alpha}(t),X_{S,\beta}(0)]\>_0=\<[J_{S,\alpha}(0),X_{S,\beta}(-t)]\>_0$ resulting from time translation invariance, performing the integration by parts yields
\begin{align}\label{eq_app:conductivity_[J,X]}
\sigma_{\alpha\beta}^{(S)}(\omega)=-\frac{1}{\omega^+}\<[J_{S,\alpha}(0),X_{S,\beta}(0)]\>_0+\frac{i}{\omega^+}\chi_{\alpha\beta}(\omega),
\end{align}
where
\begin{align}\label{eq_app:chi}
\chi_{\alpha\beta}(\omega)=-i\int_{-\infty}^\infty\!\!dt\,e^{i\omega^+ t}\theta(t)\<[J_{S,\alpha}(t),J_{S,\beta}(0)]\>_0
\end{align}
is the retarded response function for the spin current.
Using Eqs.~(\ref{eq_app:[r,p]}) and (\ref{eq_app:J_S(t)}), we obtain
\begin{align}\label{eq_app:[J_S,X_S]}
[J_{S,\alpha}(0),X_{S,\beta}(0)]=-i\delta_{\alpha\beta}\sum_{s_z}\frac{s_z^2N_{s_z}}{m}.
\end{align}
Substituting this into Eq.~(\ref{eq_app:conductivity_[J,X]}), we finally find Eq.~\eqref{eq:conductivity1}:
\begin{align}\label{eq_app:conductivity1}
\sigma_{\alpha\beta}^{(S)}(\omega)=\frac{i}{\omega^+}\left(\delta_{\alpha\beta}\sum_{s_z}\frac{s_z^2N_{s_z}}{m}+\chi_{\alpha\beta}(\omega)\right).
\end{align}

We next turn to the $f$-sum rule [Eq.~\eqref{eq:f-sum}], which is the exact constraint on the integral of $\sigma_{\alpha\beta}^{(S)}(\omega)$ over $\omega$.
The causality condition in the retarded function $\chi_{\alpha\beta}(\omega)$ leads to the Kramers-Kronig relation:
\begin{align}\label{eq_app:Kramers-Kronig}
\Re\,\chi_{\alpha\beta}(\omega)=\mathcal{P}\int_{-\infty}^\infty\frac{d\omega'}{\pi}\frac{\Im\,\chi_{\alpha\beta}(\omega')}{\omega'-\omega},
\end{align}
where $\mathcal{P}$ denotes the Cauchy principal value.
Using this relation, we can obtain the $f$-sum rule [Eq.~\eqref{eq:f-sum}]:
\begin{align}\label{eq_app:f-sum}
\int_{-\infty}^{\infty}\!\!\frac{d\omega}{\pi}\Re\,\sigma_{\alpha\beta}^{(S)}(\omega)=\delta_{\alpha\beta}\sum_{s_z}\frac{s_z^2 N_{s_z}}{m}.
\end{align}

\section{\label{appendix:S=1/2}Spin-1/2 superfluid Fermi gas}
Here we compute the optical spin conductivity for a spin-1/2 Fermi superfluid at zero temperature within the BCS-Leggett mean-field theory~\cite{Eagles:1969,Leggett:1980}.
In the mean-field theory, Eq.~(\ref{eq:H_fermion}) reduces to
\begin{align}\label{eq_app:H_BCS}
K_\mathrm{BCS}=E_\mathrm{GS}+\sum_{\k,\sigma}E_{\k,{\rm F}}\gamma_{\k,\sigma}^\+\gamma_{\k,\sigma},
\end{align}
where $E_{\k,{\rm F}}=\sqrt{(\varepsilon_\k-\mu)^2+\Delta^2}$ is the quasiparticle energy with the superfluid order parameter $\Delta$.
Since the ground state energy $E_\mathrm{GS}$ does not contribute to spin transport, we do not provide its explicit form.
The creation and annihilation operators $\gamma_{\k,\sigma}^\+,\,\gamma_{\k,\sigma}$ of quasiparticles are given by the Bogoliubov transformation:
\begin{align}\label{eq_app:Bogoliubov_fermion}
\begin{pmatrix}
\gamma_{\k,\up}^\+\\
\gamma_{-\k,\down}
\end{pmatrix}
=
\begin{pmatrix}
u_{\k,{\rm F}}&-v_{\k,{\rm F}}\\
v_{\k,{\rm F}}&u_{\k,{\rm F}}
\end{pmatrix}
\begin{pmatrix}
c_{\k,\up}^\+\\
c_{-\k,\down}
\end{pmatrix}
\end{align}
with
\begin{align}\label{eq_app:BCS_uv}
u_{\k,{\rm F}}=\sqrt{\frac{1}{2}\left(1+\frac{\varepsilon_\k-\mu}{E_{\k,{\rm F}}}\right)},\quad v_{\k,{\rm F}}=\sqrt{\frac{1}{2}\left(1-\frac{\varepsilon_\k-\mu}{E_{\k,{\rm F}}}\right)}.
\end{align}
The operators $\gamma_{\k,\sigma}^\+,\,\gamma_{\k,\sigma}$ satisfy the following anticommutation relations:
\begin{subequations}\label{eq_app:gamma,gamma}
\begin{align}
\{\gamma_{\k,\sigma},\gamma_{\k',\sigma'}\}&=\{\gamma_{\k,\sigma}^\+,\gamma_{\k',\sigma'}^\+\}=0,\\
\{\gamma_{\k,\sigma},\gamma_{\k',\sigma'}^\+\}&=\delta_{\k\k'}\delta_{\sigma\sigma'}.
\end{align}
\end{subequations}

In the mean field approximation, $\Delta$ and $\mu$ for given $a$ and $N=N_\up+N_\down$ are determined by self-consistently solving the following gap and particle number equations:
\begin{align}
-\frac{m}{4\pi a}&=\frac{1}{\Omega}\sum_\k\left(\frac{1}{2E_{\k,{\rm F}}}-\frac{1}{2\varepsilon_\k}\right),\\
\label{eq_app:number}
N&=\sum_\k\left(1-\frac{\varepsilon_\k-\mu}{E_{\k,{\rm F}}}\right).
\end{align}

\subsection{Current correlation function}
Here, we calculate the correlation function $\chi_{\alpha\beta}(\omega)$ in Eq.~(\ref{eq_app:chi}) for the superfluid Fermi gas.
In the second quantization formalism, $\bm{J}_S$ in Eq.~(\ref{eq_app:J_S(t)}) is rewritten as
\begin{align}
\bm{J}_S=\sum_\k\frac{\k}{2m}(c_{\k,\up}^\+c_{\k,\up}-c_{\k,\down}^\+c_{\k,\down}).
\end{align}
Substituting the inverse of the Bogoliubov transformation in Eq.~(\ref{eq_app:Bogoliubov_fermion}) into this yields
\begin{align}
\bm{J}_S&=\sum_\k\frac{\k}{2m}\left[(u_{\k,{\rm F}}^2-v_{\k,{\rm F}}^2)(\gamma_{\k,\up}^\+\gamma_{\k,\up}+\gamma_{-\k,\down}^\+\gamma_{-\k,\down})
\right.\nonumber\\
&\quad\left.+2v_{\k,{\rm F}}^2
+2u_{\k,{\rm F}}v_{\k,{\rm F}}(\gamma_{\k,\up}^\+\gamma_{-\k,\down}^\++\gamma_{-\k,\down}\gamma_{\k,\up})\right].
\end{align}
In the Heisenberg picture, $\bm{J}_S(t)=U_\mathrm{BCS}^\+(t)\bm{J}_SU_\mathrm{BCS}(t)$ with $U_\mathrm{BCS}(t)=\exp(-iK_\mathrm{BCS}\,t)$ reads
\begin{align}\label{eq_app:J_S(t)_fermion}
\bm{J}_S(t)&=\sum_\k\frac{\k}{2m}\Bigl[(u_{\k,{\rm F}}^2-v_{\k,{\rm F}}^2)(\gamma_{\k,\up}^\+\gamma_{\k,\up}+\gamma_{-\k,\down}^\+\gamma_{-\k,\down})\nonumber\\
&\quad+2u_{\k,{\rm F}}v_{\k,{\rm F}}(e^{2iE_{\k,{\rm F}}t}\gamma_{\k,\up}^\+\gamma_{-\k,\down}^\++e^{-2iE_{\k,{\rm F}}t}\gamma_{-\k,\down}\gamma_{\k,\up})\nonumber\\
&\quad+2v_{\k,{\rm F}}^2\Bigr],
\end{align}
where $\gamma_{\k,\sigma}(t)=U_\mathrm{BCS}^\+(t)\gamma_{\k,\sigma}U_\mathrm{BCS}(t)=\gamma_{\k,\sigma}e^{-iE_{\k,{\rm F}}t}$ was used.

Let us now evaluate the correlation function in Eq.~(\ref{eq_app:chi}) at zero temperature.
Using Eqs.~(\ref{eq_app:BCS_uv}), (\ref{eq_app:gamma,gamma}), and (\ref{eq_app:J_S(t)_fermion}) as well as $\<\gamma_{\k,\sigma}^\+\gamma_{\k,\sigma}\>_0=0$ at zero temperature, the expectation value in Eq.~(\ref{eq_app:chi}) is
\begin{align}
\<[J_{S,\alpha}(t),J_{S,\beta}(0)]\>_0=\sum_{\k}\frac{\Delta^2k_\alpha k_{\beta}}{4m^2E_{\k,{\rm F}}^2}\left(e^{-2iE_{\k,{\rm F}}t}-e^{2iE_{\k,{\rm F}}t}\right).
\end{align}
Therefore, the correlation function for the superfluid Fermi gas is found to be
\begin{align}\label{eq_app:chi_fermions}
\chi_{\alpha\beta}(\omega)=\delta_{\alpha\beta}\sum_\k\frac{\Delta^2k_\alpha^2}{4m^2E_{\k,{\rm F}}^2}
\left(\frac{1}{\omega^+-2E_{\k,{\rm F}}}-\frac{1}{\omega^++2E_{\k,{\rm F}}}\right).
\end{align}

\subsection{Spin conductivity}
Let us evaluate the real part of the optical spin conductivity.
Equation~(\ref{eq:conductivity1}) provides
\begin{align}\label{eq_app:conductivity_real}
\Re\,\sigma_{xx}^{(S)}(\omega)=\mathcal{D}_S\delta(\omega)-\frac{1}{\omega}\Im\,\chi_{xx}(\omega),
\end{align}
where $\mathcal{D}_S=\pi\left[N/4m+\Re\,\chi_{xx}(0)\right]$ is the spin Drude weight.
Using Eqs.~(\ref{eq_app:number}) and (\ref{eq_app:chi_fermions}), replacing $\sum_\k\to\Omega\int d^3\k/(2\pi)^3$, and performing the integration over $k=|\k|$ by parts, we can find $\mathcal{D}_S=0$ in this case.
By substituting Eq.~(\ref{eq_app:chi_fermions}) into the second term in Eq.~(\ref{eq_app:conductivity_real}), the real part of $\sigma_{xx}^{(S)}(\omega)$ is given by
\begin{align}\label{eq_app:conductivity_fermions_1}
\Re\,\sigma_{xx}^{(S)}(\omega)&=\sum_\k\frac{\pi \Delta^2k_x^2}{m^2|\omega|^3}\delta(|\omega|-2E_{\k,{\rm F}}).
\end{align}
Using Eq.~(\ref{eq_app:number}), we can straightforwardly confirm that this spin conductivity satisfies the $f$-sum rule in Eq.~(\ref{eq:f-sum}).

\section{\label{appendix:S=1}Spin-1 polar Bose-Einstein condensate}
Here we compute the optical spin conductivity for a spin-1 Bose-Einstein condensate (BEC) at zero temperature within the Bogoliubov theory~\cite{kawaguchi}.
In the polar phase, the condensate is characterized by $\langle a_{\k=\mathbf{0},s_z} \rangle=\sqrt{n_0}\delta_{s_z0}$
with the condensate fraction $n_0$ and is stabilized in the plane of $(q,n_0c_1)$ satisfying $q+n_0c_1>n_0|c_1|$.
By using the Bogoliubov theory, where the effect of $a_{\k\ne\mathbf{0}}$ on $K$ [Eq.~(\ref{eq:H_boson})] is incorporated up to quadratic order, $K$ reduces to~\cite{Uchino:2010}
\begin{align}
K_\mathrm{Bog}&=E_\mathrm{GS}+\sum_{\k\neq\0}\left[E_{\k,d}\beta^{\dagger}_{\k,d}\beta_{\k,d}\right.\cr
&\quad\left.+E_{\k,s}(\beta^{\dagger}_{\k,s_x}\beta_{\k,s_x}+\beta^{\dagger}_{\k,s_y}\beta_{\k,s_y})\right].
\end{align}
Since the ground-state energy $E_\mathrm{GS}$ does not contribute to spin transport as in the case of the spin-$1/2$ superfluid Fermi gas, we do not show its explicit form.
The quasiparticle energies in the density ($d$) and spin ($s$) channels are given by $E_{\k,d}=\sqrt{\varepsilon_\k(\varepsilon_\k+2n_0c_0)}$ and $E_{\k,s}=\sqrt{(\varepsilon_\k+q)(\varepsilon_\k+q+2n_0c_1)}$, respectively.
The operators $\beta_{\k,d}$, $\beta_{\k,s_x}$, and $\beta_{\k,s_y}$ denote the annihilation operators of quasiparticles, which are related to 
\begin{subequations}\label{eq_app:unitary_transform}
\begin{align}
b_{\k ,d}&=a_{\k,0},\\
b_{\k,s_x}&=\frac{1}{\sqrt{2}}(a_{\k,1}+a_{\k,-1}),\\
b_{\k,s_y}&=\frac{i}{\sqrt{2}}(a_{\k,1}-a_{\k,-1})
\end{align}
\end{subequations}
by the Bogoliubov transformations:
\begin{subequations}\label{eq_app:Bogoliubov_boson}
\begin{align}
b_{\k,d}&=u_{\k,d}\beta_{\k,d}-v_{\k,d}\beta^{\dagger}_{-\k,d},\\
b_{\k,s_x}&=u_{\k,s}\beta_{\k,s_x}-v_{\k,s}\beta^{\dagger}_{-\k,s_x},\\
b_{\k,s_y}&=u_{\k,s}\beta_{\k,s_y}-v_{\k,s}\beta^{\dagger}_{-\k,s_y}
\end{align}
\end{subequations}
with
\begin{subequations}\label{eq_app:Bogoliubov_uv}
\begin{align}
u_{\k,d}&=\sqrt{\frac{\varepsilon_\k+n_0c_0+E_{\k,d} } {2E_{\k,d}}},\\
v_{\k,d}&=\sqrt{\frac{\varepsilon_\k+n_0c_0-E_{\k,d} } {2E_{\k,d}}},\\
u_{\k,s}&=\sqrt{\frac{\varepsilon_\k+q +n_0c_1+E_{\k,s} } {2E_{\k,s}}},\\
v_{\k,s}&=\sqrt{\frac{\varepsilon_\k+q+n_0c_1-E_{\k,s} } {2E_{\k,s}}}.
\end{align}
\end{subequations}
Since the density channel does not contribute to spin transport, we below consider the spin channels.
The annihilation and creation operators of quasiparticles in the spin channels satisfy the following commutation relations:
\begin{subequations}\label{eq_app:beta,beta}
\begin{align}
[\beta_{\k,s_j},\beta_{\k,s_{j'}}]&=[\beta_{\k,s_{j}}^\+,\beta_{\k,s_{j'}}^\+]=0,\\
[\beta_{\k,s_j},\beta_{\k,s_{j'}}^\+]&=\delta_{\k\k'}\delta_{s_js_{j'}}
\end{align}
\end{subequations}
with $j,j'\in\{x,y\}$.

\subsection{Current correlation function}
Here, we calculate the correlation function $\chi_{\alpha\beta}(\omega)$ in Eq.~(\ref{eq_app:chi}) for the spinor BEC in the polar phase.
In the second quantization formalism, $\bm{J}_S$ in Eq.~(\ref{eq_app:J_S(t)}) is rewritten as
\begin{align}
\bm{J}_S&=\sum_\k\frac{\k}{m}(a_{\k,1}^\+a_{\k,1}-a_{\k,-1}^\+a_{\k,-1})\cr
&=-i\sum_{\k}\frac{\k}{m}( b^{\dagger}_{\k,s_x }b_{\k,s_y}-b^{\dagger}_{\k,s_y}b_{\k,s_x}),
\end{align}
where Eqs.~(\ref{eq_app:unitary_transform}) were used.
Substituting Eqs.~(\ref{eq_app:Bogoliubov_boson}) into this and using Eqs.~(\ref{eq_app:beta,beta}), we obtain
\begin{align}
\bm{J}_S&=-i\sum_{\k}\frac{\k}{m}\left[(u_{\k,s}^2+v_{\k,s}^2)(\beta^{\dagger}_{\k,s_x}\beta_{\k,s_y}-\beta^{\dagger}_{\k,s_y}\beta_{\k,s_x})\right.\nonumber\\
&\quad\left.-2u_{\k,s}v_{\k,s}(\beta^{\dagger}_{\k,s_x}\beta^{\dagger}_{-\k,s_y}-\beta_{\k,s_x}\beta_{-\k,s_y})\right].
\end{align}
In the Heisenberg picture, $\bm{J}_S(t)=U_\mathrm{Bog}^\+(t)\bm{J}_SU_\mathrm{Bog}(t)$ with $U_\mathrm{Bog}(t)=\exp(-iK_\mathrm{Bog}\,t)$ reads
\begin{align}\label{eq_app:J_S(t)_boson}
\bm{J}_S(t)&=\sum_{\k}\frac{\k}{m}\left[-i(u_{\k,s}^2+v_{\k,s}^2)\beta^{\dagger}_{\k,s_x}\beta_{\k,s_y}\right.\nonumber\\
&\quad\left.+2i\,u_{\k,s}v_{\k,s}e^{2iE_{\k,s}t}\beta^{\dagger}_{\k,s_x}\beta^{\dagger}_{-\k,s_y}+\mathrm{h.c.}\right],
\end{align}
where $\beta_{\k,s_j}(t)=U_\mathrm{Bog}^\+(t)\beta_{\k,s_j}U_\mathrm{Bog}(t)=\beta_{\k,s_j}e^{-iE_{\k,s}t}$ was used.

Let us now evaluate the correlation function in Eq.~(\ref{eq_app:chi}) at zero temperature.
Using Eqs.~(\ref{eq_app:Bogoliubov_uv}), (\ref{eq_app:beta,beta}), and (\ref{eq_app:J_S(t)_boson}) as well as $\<\beta_{\k,s_j}^\+\beta_{\k,s_j}\>_0=0$ at zero temperature, the expectation value in Eq.~(\ref{eq_app:chi}) is
\begin{align}
\<[J_{S,\alpha}(t),J_{S,\beta}(0)]\>_0=\sum_{\k}\frac{n_0^2c_1^2k_{\alpha}k_{\beta}}{m^2E_{\k,s}^2}\left(e^{-2iE_{\k,s}t}-e^{2iE_{\k,s}t}\right).
\end{align}
Therefore, the correlation function for the polar BEC is found to be
\begin{align}\label{eq_app:chi_bosons}
\chi_{\alpha\beta}(\omega)=\delta_{\alpha\beta}\sum_\k\frac{n_0^2c_1^2k_\alpha^2}{m^2E_{\bm{k},s}^2}\left(\frac{1}{\omega^+-2E_{\bm{k},s}}-\frac{1}{\omega^++2E_{\bm{k},s}}\right).
\end{align}

\subsection{Spin conductivity}
Let us now evaluate the real part of the optical spin conductivity.
The real part is given as the form of Eq.~(\ref{eq_app:conductivity_real}), where the spin Drude weight $\mathcal{D}_S=\pi\left[(N_1+N_{-1})/m+\Re\,\chi_{xx}(0)\right]$ in this case involves the quantum depletion $N_1+N_{-1}$ in the spin channels.
Using $N_1+N_{-1}=\sum_\k2v_{\k,s}^2$~\cite{Uchino:2010} and Eq.~(\ref{eq_app:chi_bosons}), we can see $\mathcal{D}_S=0$ in a similar way as for the superfluid Fermi gas.
The real part thus reads
\begin{align}\label{eq_app:conductivity_bosons_1}
\Re\,\sigma_{xx}^{(S)}(\omega)=\sum_\k\frac{4\pi n_0^2c_1^2k_x^2}{m^2|\omega|^3}\delta(|\omega|-2E_{\bm{k},s}).
\end{align}
We can straightforwardly confirm that this spin conductivity satisfies the $f$-sum rule in Eq.~(\ref{eq:f-sum}).
We note that the $f$ sum in this case is equivalent to the quantum depletion in the spin channel~\cite{Uchino:2010} while that in the $S=1/2$ case provides the total particle number.

\section{\label{appendix:TL}Tomonaga-Luttinger liquid}
Here we consider spin-1/2 one-dimensional quantum fluids where the low-energy description based on Tomonaga-Luttinger (TL) liquids is reasonable.
In addition to the Hamiltonian $H=H_{C}+H_{S}$ [see Eqs.~\eqref{eq:H_C/S}], one can bosonize physical quantities in this case.
For instance, the local current operator  is expressed as
\begin{eqnarray}
j_{C(S)}(x,t)=\sqrt{2}u_{C(S)}K_{C(S)}\Pi_{C(S)}(x,t).
\end{eqnarray}
Owing to the spin-charge separation and the formal similarity of the Hamiltonian between charge and spin sectors, as in the case of the charge conductivity~\cite{giamarchi},
one can  obtain the spin conductivity expression as
\begin{eqnarray}
\sigma^{(S)}(\omega)=\frac{i}{\omega}\left(\frac{2u_SK_S\Omega}{\pi}+\chi(\omega)\right),
\end{eqnarray}
where $\Omega$ is the one-dimensional volume and the retarded current-current correlation function $\chi(\omega)$ is  given by Eq.~(\ref{eq_app:chi}) with $J_S(t)=\int dx\,j_S(x,t)$.
In order to obtain the finite frequency dependence, we rewrite the conductivity as 
\begin{eqnarray}\label{eq_app:memory}
\sigma^{(S)}(\omega)=\frac{2iu_SK_S\Omega}{\pi}\frac{1}{\omega+M(\omega)},
\end{eqnarray}
where $M(\omega)$ is called the memory function.
One can perturbatively calculate $M(\omega)$, provided that $g$ is small.
By performing the similar calculation with charge transport~\cite{giamarchi1991,giamarchi1992}, we obtain
\begin{align}
M(\omega)&\approx
\frac{g^2K_S}{\pi^3\alpha^2}\sin(2\pi K_S)\Gamma^2(1-2K_S)\frac{e^{-i\pi(2K_S-1)}}{\omega}\nonumber\\
&\quad\times
\left(\frac{\alpha \omega}{2u_S}\right)^{4K_S-2},
\end{align}
where $\Gamma(z)$ is the gamma function and we assumed the zero temperature.
For $K_{S}\ge1$ that corresponds to cases of repulsive interactions,
 $\cos(\sqrt{8}\phi_S)$ becomes irrelevant and therefore the TL liquid is realized.
In this case, $M(\omega)$ is negligible compared with $\omega$  at low frequencies. Thus, by using Eq.~\eqref{eq_app:memory}, we find $\text{Re}[\sigma^{(S)}(\omega)]\sim \omega^{4K_S-5}$.

\section{\label{sec:single_frequency}Spin dynamics under single-frequency driving}
Here, we derive Eq.~\eqref{eq:X_S(t)}, which allows us to experimentally extract $\sigma_{\alpha\beta}^{(S)}(\omega)$ from spin dynamics driven by the perturbation [Eq.~(\ref{eq:deltaH(t)})].
We consider one of the simplest perturbations, i.e., the single-frequency driving $f_\beta(t)=F_\beta\cos(\omega_0t)$.
In this case, substituting the Fourier transform of $f_\beta(t)$ into Eq.~(\ref{eq:conductivity}) and performing the inverse Fourier transform of $\<\tilde{J}_{S,\alpha}(\omega)\>$ yields
\begin{align}\label{eq_app:oscillating_J_S(t)}
\frac{\<J_{S,\alpha}(t)\>}{F_\beta}
=\Re\,\sigma_{\alpha\beta}^{(S)}(\omega_0)\cos(\omega_0t)+\Im\,\sigma_{\alpha\beta}^{(S)}(\omega_0)\sin(\omega_0t),
\end{align}
where $\sigma_{\alpha\beta}^{(S)}(-\omega_0)=[\sigma_{\alpha\beta}^{(S)}(\omega_0)]^*$ resulting from the hermiticity of $J_{S,\alpha}$ and $X_{S,\beta}$ in Eq.~(\ref{eq_app:conductivity0}) was used.
Because of $J_{S,\alpha}(t)=\frac{d}{dt}X_{S,\alpha}(t)$,
$\<\delta X_{S,\alpha}(t)\>\equiv\<X_{S,\alpha}(t)\>-\<X_{S,\alpha}\>_0$ shows an oscillating behavior [Eq.~\eqref{eq:X_S(t)}]:
\begin{align}\label{eq_app:oscillating_X_S(t)}
&\frac{\<\delta X_{S,\alpha}(t)\>}{F_\beta}\nonumber
\\
&=-\frac{\Im\,\sigma_{\alpha\beta}^{(S)}(\omega_0)}{\omega_0}\cos(\omega_0t)
+\frac{\Re\,\sigma_{\alpha\beta}^{(S)}(\omega_0)}{\omega_0}\sin(\omega_0t).
\end{align}
Therefore, $\sigma_{\alpha\beta}^{(S)}(\omega=\omega_0)$ can be experimentally determined by measuring the oscillation of $\<J_{S,\alpha}(t)\>$ or $\<X_{S,\alpha}(t)\>$ under the perturbation $f_\beta(t)=F_\beta\cos(\omega_0t)$.

\section{\label{appendix:spin_mass}Measurement scheme in the presence of a mass current}
In this Appendix, we extend our proposal to the cases where both mass and spin currents are induced by a time-dependent force.
Such a situation is realized when a magnetic-field gradient is applied to $^{40}$K atoms~\cite{Jotzu:2015}.
We show that this type of perturbation is also available to experimentally extract the optical spin conductivity of harmonically trapped or homogeneous gases with two internal degrees of freedom.
For these systems the center-of-mass motion is not affected by the interparticle interactions~\cite{Kohn:1961,Brey:1989,Li:1991}.
We note that the discussion below is not limited to spin-1/2 Fermi gases and holds for two-component Bose gases if two species are referred to as spin-up and spin-down states.

We start with the following Hamiltonian in the presence of a harmonic trapping potential:
$\H(t)=H+\delta H_\beta'(t)$, where the nonperturbative and perturbative terms are given by
\begin{align}\label{eq_app:H_asymmetric}
H&=\sum_{\sigma=\up,\down}\int d\r\,\left(\frac{|{\bm{\nabla}\psi_\sigma(\r)|}^2}{2m}+\sum_{\alpha=x,y,z}\frac12m\omega_\alpha^2r_\alpha^2 n_\sigma(\r)\right)\nonumber\\
&\quad+\sum_{\sigma,\sigma'}H_\mathrm{int}^{\sigma\sigma'},\\
\delta H_\beta'(t)
&=-\sum_{\sigma=\up,\down}\gamma_\sigma f_\beta(t)X_{\sigma\beta},
\end{align}
respectively, and $X_{\sigma\beta}=\int d\r\,r_\beta n_\sigma(\r)$.
Here, $\psi_\sigma(\r)$ and $n_\sigma(\r)=\psi_\sigma^\+(\r)\psi_\sigma(\r)$ are the field and particle number operators of spin-$\sigma$ particles, respectively, and $\omega_\alpha$ is a trapping frequency.
The interaction terms $H_\mathrm{int}=\sum_{\sigma,\sigma'}H_\mathrm{int}^{\sigma\sigma'}$ have the form of
\begin{align}\label{eq_app:H_int}
H_\mathrm{int}^{\sigma\sigma'}=\frac{1}{2}\int d\r d\r'\,\psi_\sigma^\+(\r)\psi_{\sigma'}^\+(\r')U_{\sigma\sigma'}(\r-\r')\psi_{\sigma'}(\r')\psi_\sigma(\r)
\end{align}
with the interaction potentials $U_{\sigma\sigma'}(\r-\r')$ and thus satisfies $[H_\mathrm{int},S_z(\r)]=0$ with $S_z(\r)=[n_\up(\r)-n_\down(\r)]/2$.
The parameter $\gamma_\sigma$ characterizes the strength of the external force to kick spin-$\sigma$ particles.
The perturbative term can be separated into mass and spin components as
\begin{align}\label{eq_app:deltaH(t)_asymmetric}
\delta H_\beta'(t)&=-f_\beta(t)(\gamma_M X_{M,\beta}+\gamma_S X_{S,\beta}),
\end{align}
where $\gamma_M=(\gamma_\up+\gamma_\down)/2$, $\gamma_S=\gamma_\up-\gamma_\down$, $X_{M,\beta}=X_{\up\beta}+X_{\down\beta}$, and $X_{S,\beta}=(X_{\up\beta}-X_{\down\beta})/2$.
This definition of $\X_S$ is consistent with Eq.~(\ref{eq_app:deltaH(t)}) in the first quantization, while $\X_M$ describes the center-of-mass coordinate~\footnote{Strictly speaking, $\X_M$ is related to the center-of-mass coordinate ${\bm R}_M$ as $\X_M=N{\bm R}_M$.}.
For $\gamma_\up=\gamma_\down$ ($\gamma_S=0$), a pure mass current is driven, while for $\gamma_\up=-\gamma_\down$ ($\gamma_M=0$) a pure spin current is driven.
Hereafter, we focus on $\gamma_\up\neq|\gamma_\down|$, where both mass and spin currents flow.
By taking the limit of $\omega_\alpha\to0$ below, we can also obtain results for the homogeneous case.

The responses of the mass current $\J_M(t)=\frac{d\X_M(t)}{dt}$ and spin current $\J_S(t)=\frac{d\X_S(t)}{dt}$ to the perturbation have the following forms in frequency space:
\begin{align}
\frac{\<\tilde{J}_{M,\alpha}(\omega)\>}{\tilde{f}_\beta(\omega)}&=\gamma_M\sigma_{\alpha\beta}^{MM}(\omega)+\gamma_S\sigma_{\alpha\beta}^{MS}(\omega),\\
\label{eq_app:J_S_asymmetric}
\frac{\<\tilde{J}_{S,\alpha}(\omega)\>}{\tilde{f}_\beta(\omega)}&=\gamma_M\sigma_{\alpha\beta}^{SM}(\omega)+\gamma_S\sigma_{\alpha\beta}^{SS}(\omega),
\end{align}
where $\tilde{J}_{M,\alpha}(\omega)$, $\tilde{J}_{S,\alpha}(\omega)$, and $\tilde{f}_\beta(\omega)$ are Fourier transforms of $J_{M,\alpha}(t)$, $J_{S,\alpha}(t)$, and $f_\beta(t)$, respectively, and
\begin{align}\label{eq_app:generalized_conductivity}
\sigma_{\alpha\beta}^{\mathrm{a}\mathrm{b}}(\omega)
=i\int_0^\infty\!\!dt\,e^{i\omega^+ t}
\<[J_{\mathrm{a},\alpha}(t),X_{\mathrm{b},\beta}(0)]\>_0
\end{align}
with $\mathrm{a},\mathrm{b}\in\{M,S\}$ is the generalized optical conductivity.
The mass and spin conductivities are given by $\sigma_{\alpha\beta}^{(M)}(\omega)=\sigma_{\alpha\beta}^{MM}(\omega)$ and $\sigma_{\alpha\beta}^{(S)}(\omega)=\sigma_{\alpha\beta}^{SS}(\omega)$, respectively.
In Eq.~(\ref{eq_app:generalized_conductivity}), the nonperturbative Hamiltonian [Eq.~(\ref{eq_app:H_asymmetric})] governs the time evolution of operators.
In the case of the harmonic trap, the equation of motion of $\X_M(t)$ is independent of the interaction term $\left[\frac{d^2 X_{M,\alpha}(t)}{dt^2}=-\omega_\alpha^2X_{M,\alpha}(t)\right]$ and can be easily solved:
\begin{align}\label{eq_app:X_M(t)_trivial}
X_{M,\alpha}(t)
&=\cos(\omega_\alpha t)X_{M,\alpha}+\frac{\sin(\omega_\alpha t)}{\omega_\alpha}J_{M,\alpha}.
\end{align}
By substituting this into Eq.~(\ref{eq_app:generalized_conductivity}) and using canonical anticommutation (or commutation) relations of field operators, we can see that the conductivities including mass degrees of freedom have the following trivial forms:
\begin{align}
\sigma_{\alpha\beta}^{(M)}(\omega)&=\delta_{\alpha\beta}N\sigma_\alpha^0(\omega),\\
\label{eq_app:cross_conductivity}
\sigma_{\alpha\beta}^{MS}(\omega)&=
\sigma_{\alpha\beta}^{SM}(\omega)=\delta_{\alpha\beta}\frac{N_\up-N_\down}{2}\sigma_\alpha^0(\omega),
\end{align}
where $N_\sigma=\int d\r\,\<\psi_\sigma^\+(\r)\psi_\sigma(\r)\>_0$, $N=N_\up+N_\down$, and
\begin{align}
\sigma_\alpha^0(\omega)&=\frac{i}{2m}\left(\frac{1}{\omega^+-\omega_\alpha}+\frac{1}{\omega^++\omega_\alpha}\right).
\end{align}
This explicit form of $\sigma_{\alpha\beta}^{SM}(\omega)$ allows us to experimentally extract the optical spin conductivity by measuring $\<\X_{S}(t)\>$ or $\<\J_{S}(t)\>$ even in the presence of the mass current.
Indeed, we can find in a similar way as in Sec.~\ref{sec:single_frequency} that $\<X_{S,\alpha}(t)\>$ under the single-frequency driving $f_\beta(\omega)=F_\beta\cos(\omega_0t)$ shows an oscillating behavior:
\begin{align}\label{eq_app:X_S_asymmetric}
\frac{\<\delta X_{S,\alpha}(t)\>}{F_\beta}&=-\frac{\Im\,[\gamma_S\sigma_{\alpha\beta}^{(S)}(\omega_0)+\gamma_M\sigma_{\alpha\beta}^{SM}(\omega_0)]}{\omega_0}\cos(\omega_0t)\nonumber\\
&\quad+\frac{\Re\,[\gamma_S\sigma_{\alpha\beta}^{(S)}(\omega_0)+\gamma_M\sigma_{\alpha\beta}^{SM}(\omega_0)]}{\omega_0}\sin(\omega_0t).
\end{align}
For $\gamma_M=0$ and $\gamma_S=1$, this is consistent with Eq.~(\ref{eq:X_S(t)}).
We can experimentally determine $\sigma_{\alpha\beta}^{(S)}(\omega)$ from the oscillation of $\<X_{S,\alpha}(t)\>$ in a similar way as in the case where a pure spin current is generated.
In particular, when spin is balanced ($N_\up=N_\down$), the cross conductivity $\sigma_{\alpha\beta}^{SM}(\omega)$ in Eq.~\eqref{eq_app:cross_conductivity} vanishes, so that Eq.~(\ref{eq_app:X_S_asymmetric}) becomes equivalent to Eq.~(\ref{eq:X_S(t)}) in the case where a pure spin current is induced.
We note that the above discussion relies on the separation of the driving force into mass and spin sectors [Eq.~(\ref{eq_app:deltaH(t)_asymmetric})], which is specific to two-component systems, as well as the trivial motion of center of mass [Eq.~(\ref{eq_app:X_M(t)_trivial})] in the cases of a harmonically trapped gas without optical lattice or of a homogeneous gas.

\section{\label{appendix:SOC}Extension to systems without spin conservation}
This Appendix is devoted to extending our scheme of the detection of the optical spin conductivity to systems  without the spin conservation.
In particular, we focus on two-component gases in the presence of spin-orbit and Rabi couplings, which are realized with ultracold atoms~\cite{Lin:2011,Wang:2012,Cheuk:2012}.
The Hamiltonian in {the} second quantization is given by
\begin{align}
H_\mathrm{SOC}
&=\int\!d\r\,\Psi^\+(\r)h(\r)\Psi(\r)+H_\mathrm{int},\\
h(\r)&=-\frac{1}{2m}\D^2+V(\r)\mathbbm{1}_S+\frac{\delta}{2}\,\hat{\sigma}_z+\frac{\Omega_\mathrm{R}}{2}\,\hat{\sigma}_x,
\end{align}
where $\Psi(\r)=(\psi_\up(\r),\psi_\down(\r))^\mathrm{T}$, $\D={\bm{\nabla}}\mathbbm{1}_S-i\k_r\hat{\sigma}_z$, $\mathbbm{1}_S=\mathrm{diag}(1,1)$, and $\hat{\sigma}_{x,y,z}$ are Pauli matrices.
The spin-orbit coupling characterized by $\k_r$ can be interpreted as an equal weight combination of Rashba-type and Dresselhaus-type spin-orbit couplings~\cite{Lin:2011,Wang:2012,Cheuk:2012}, $\delta$ is a Zeeman detuning, $\Omega_\mathrm{R}$ is a Rabi coupling, and $V(\r)$ is a trapping potential.
The interaction term $H_\mathrm{int}=\sum_{\sigma\sigma'}H_\mathrm{int}^{\sigma\sigma'}$ given by Eq.~\eqref{eq_app:H_int} satisfies $[S_z(\r),H_\mathrm{int}]=0$, where the spin density operators are defined as $S_{\hat{a}}(\r)\equiv\Psi^\+(\r)\hat{\sigma}_{\hat{a}}\Psi(\r)/2$ with $\hat{a}=x,y,z$.
In the presence of the spin-orbit coupling, the operator of the spin current density involves $\D$ as in the case of the mass current with a vector potential: ${\bm{j}}_S(\r)=\frac{1}{4m}\left[\Psi^\+(\r)\{-i\D\hat{\sigma}_z\Psi(\r)\}+\{-i\D\hat{\sigma}_z\Psi(\r)\}^\+\Psi(\r)\right]$.
Therefore, the spin current operator reads
\begin{align}\label{eq_app:J_S_SOC}
\J_S=\int\!d\r\,{\bm{j}}_S(\r)=\int\!d\r\,\frac{-i}{2m}\Psi^\+(\r)\D\hat{\sigma}_z\Psi(\r).
\end{align}

We now turn to how to experimentally extract the optical spin conductivity defined by~\footnote{In Appendix~\ref{appendix:SOC}, $\sigma_{\alpha\beta}^{(S)}(\omega)$ is defined by Eq.~(\ref{eq_app:conductivity_SOC}) so as to include information on spin current correlations $\chi_{\alpha\beta}(\omega)$ and to satisfy the same $f$-sum rule as that in spin conserved cases.
Due to the source term in Eq.~(\ref{eq_app:equation_of_continuity}), $\sigma_{\alpha\beta}^{(S)}(\omega)$ is no longer equivalent to the response $\Sigma_{\alpha\beta}^{(S)}(\omega)\equiv\<\delta\tilde{\dot{X}}_{S,\alpha}^z(\omega)\>_{f_\beta^z}/\tilde{f}_\beta^z(\omega)$, which corresponds to the definition of the optical spin conductivity in spin conserved systems [recall Eq.~(\ref{eq_app:conductivity}) as well as $\J_S(t)=\dot{\X}_{S}^z(t)$ in the presence of spin conservation].
As a result, the $f$-sum rule of $\Sigma_{\alpha\beta}^{(S)}(\omega)$ is modified as $\int_{-\infty}^\infty\frac{d\omega}{\pi}\Re\,\Sigma_{\alpha\beta}^{(S)}(\omega)=\delta_{\alpha\beta}\frac{N}{4m}-\Omega\<Y_{S,\alpha\beta}^x\>_0$.
Because of $\Sigma_{\alpha\beta}^{(S)}(\omega)=-i\omega\,\Xi_{\alpha\beta}^{zz}(\omega)$, our proposed scheme allows us to measure $\Sigma_{\alpha\beta}^{(S)}(\omega)$.
In addition, the $f$ sum of $\Sigma_{\alpha\alpha}^{(S)}(\omega)$ appears in the ultrafast response of $X_{S,\alpha}^{z}(t)$ as proposed in Ref.~\cite{Carlini:2021}.
%We note that the term $\frac{N}{4m}$ in our $f$ sum is consistent with that in Eq.~(14) in Ref.~\cite{Carlini:2021}, while $-\Omega\<Y_{S,\alpha\alpha}^x\>_0$ differs.
%\del{We note that the term $\frac{N}{4m}$ in our $f$ sum is consistent with that in Eq.~(14) in Ref.~\cite{Carlini:2021}, while $-\Omega\<Y_{S,\alpha\alpha}^x\>_0$ differs.}
We note that our $f$-sum rule of $\Sigma_{\alpha\alpha}^{(S)}(\omega)$ is consistent with that in Eq.~(14) in Ref.~\cite{Carlini:2021}, where the single minimum phase with $\<n(\r)\>_0\equiv\<\Psi^\+(\r)\Psi(\r)\>_0=2\<S_x(\r)\>_0$ is focused on.
}
\begin{align}\label{eq_app:conductivity_SOC}
\sigma_{\alpha\beta}^{(S)}(\omega)
\equiv\frac{i}{\omega^+}\left(\delta_{\alpha\beta}\frac{N}{4m}+\chi_{\alpha\beta}(\omega)\right),
\end{align}
where $N=\int d\r\,\<\Psi^\+(\r)\Psi(\r)\>_0$ and the correlation function $\chi_{\alpha\beta}(\omega)=-i\int_0^\infty\!dt\,e^{i\omega^+ t}\<[J_{S,\alpha}(t),J_{S,\beta}(0)]\>_0$ now involves $\J_S$ in Eq.~(\ref{eq_app:J_S_SOC}).
Using the Kramers-Kronig relation of $\chi_{\alpha\beta}(\omega)$, we can straightforwardly show that this $\sigma_{\alpha\beta}^{(S)}(\omega)$ satisfies the same form of the $f$-sum rule as that with {the} spin conservation [Eq.~(\ref{eq:f-sum})].
In the spin-conserved case, $\sigma_{\alpha\beta}^{(S)}(\omega)$ can be determined by measuring the response of $X_{S,\alpha}(t)=\int d\r\,r_\alpha S_z(\r,t)$ to the external force coupled to $S_z(\r)$.
In the presence of the spin-orbit and Rabi couplings, $\sigma_{\alpha\beta}^{(S)}(\omega)$ can be extracted by generalizing the directions $\hat{a},\hat{b}\in\{x,y,z\}$ in {the} spin space of the measured quantity $X_{S,\alpha}(t)\to X_{S,\alpha}^{\hat{a}}(t)\equiv\int d\r\,r_\alpha S_{\hat{a}}(\r,t)$ and the perturbation $\delta H_\beta(t)\to\delta H_\beta^{\hat{b}}(t)\equiv-f_\beta^{\hat{b}}(t)X_{S,\beta}^{\hat{b}}$.
This comes from the fact that, in this case, the equation of continuity of $S_z(\r,t)$ has source terms but still gives the expression of ${\bm{\nabla}}\cdot{\bm{j}}_S(\r,t)$ in terms of measurable spin densities $S_{\hat{a}}(\r,t)$ [see Eq.~(\ref{eq_app:equation_of_continuity})].
The relation of $\sigma_{\alpha\beta}^{(S)}(\omega)$ to measurable quantities is given by
\begin{align}\label{eq_app:conductivity_SOC_result}
\sigma_{\alpha\beta}^{(S)}(\omega)
&=-i\omega\,\Xi_{\alpha\beta}^{zz}(\omega)-\Omega_\mathrm{R}\,\Xi_{\alpha\beta}^{yz}(\omega)\nonumber\\
&\quad+\frac{i\Omega_\mathrm{R}}{\omega^+}\left(-i\omega\,\Xi_{\alpha\beta}^{zy}(\omega)-\Omega_\mathrm{R}\,\Xi_{\alpha\beta}^{yy}(\omega)-\<Y_{S,\alpha\beta}^x\>_0\right),
\end{align}
where $\Xi_{\alpha\beta}^{{\hat{a}}{\hat{b}}}(\omega)$ describes the linear response of $X_{S,\alpha}^{\hat{a}}(t)$ to $f_\beta^{\hat{b}}(t)$ in {the} frequency space and the last term $\<Y_{S,\alpha\beta}^x\>_0=\int\!d\r\,r_\alpha r_\beta\<S_x(\r)\>_0$ can be determined by measuring the spin density $S_x(\r)$ at thermal equilibrium.
The quantity $\Xi_{\alpha\beta}^{{\hat{a}}{\hat{b}}}(\omega)$ can be extracted by measuring the response of $X_{S,\alpha}^{\hat{a}}(t)$ under the single-frequency driving $f_\beta^{\hat{b}}(t)=F_\beta^{\hat{b}}\cos(\omega_0t)$.
Indeed, the response takes the form of
\begin{align}\label{eq_app:<delta_X_S>_omega_0}
&\frac{\<\delta X_{S,\alpha}^{\hat{a}}(t)\>_{f_\beta^{\hat{b}}}}{F_\beta^{\hat{b}}}\nonumber\\
&=\Re\,\Xi_{\alpha\beta}^{{\hat{a}}{\hat{b}}}(\omega_0)\cos(\omega_0t)+\Im\,\Xi_{\alpha\beta}^{{\hat{a}}{\hat{b}}}(\omega_0)\sin(\omega_0t),
\end{align}
so that $\Xi_{\alpha\beta}^{{\hat{a}}{\hat{b}}}({\omega=}\omega_0)$ can be determined by observing the oscillation of $X_{S,\alpha}^{\hat{a}}(t)$.
The first and second terms in Eq.~(\ref{eq_app:conductivity_SOC_result}) can be obtained by measuring $X_{S,\alpha}^{\hat{a}=y,z}(t)$ under the external force $f_\beta^{\hat{b}=z}(t)$, while the third and forth terms by measuring $X_{S,\alpha}^{\hat{a}=y,z}(t)$ under $f_\beta^{\hat{b}=y}(t)$.
Detailed derivations of Eqs.~(\ref{eq_app:conductivity_SOC_result}) and (\ref{eq_app:<delta_X_S>_omega_0}) are presented below.

\subsection{Derivations of Eqs.~(\ref{eq_app:conductivity_SOC_result}) and (\ref{eq_app:<delta_X_S>_omega_0})}
We start with the total Hamiltonian $\H(t)$ which includes the external force $f_\beta^{\hat{b}}(t)$ along the $\beta$-direction of the coordinate space coupled to $S_{\hat{b}}(\r)$.
In the Schr\"odinger picture, we have
\begin{align}
\H(t)&=H_\mathrm{SOC}+\delta H_\beta^{\hat{b}}(t)\qquad\qquad({\hat{b}}=z,y).
\end{align}
By evaluating the Heisenberg equation of $S_{z}(\r)$, it is found that the equation of continuity of spin has source terms:
\begin{align}\label{eq_app:equation_of_continuity}
&\frac{\d S_z(\r,t)}{\d t}+{\bm{\nabla}}\cdot{\bm{j}}_S(\r,t)
\nonumber\\
&=
\begin{cases}
\Omega_\mathrm{R} S_y(\r,t)&({\hat{b}}=z\ \text{or}\ f_\beta^{\hat{b}}(t)=0),\\
\Omega_\mathrm{R} S_y(\r,t)+r_\beta f_\beta^y(t)S_x(\r,t)&({\hat{b}}=y).
\end{cases}
\end{align}
By multiplying $r_\alpha$ and integrating over $\r$, we obtain
\begin{align}\label{eq_app:dot_X_S_f^z}
\dot{X}_{S,\alpha}^z(t)
&=J_{S,\alpha}(t)+\Omega_\mathrm{R} X_{S,\alpha}^y(t)
\end{align}
for ${\hat{b}}=z\ \text{or}\ f_\beta^{\hat{b}}(t)=0$ and
\begin{align}\label{eq_app:dot_X_S_f^y}
\dot{X}_{S,\alpha}^z(t)&=J_{S,\alpha}(t)+\Omega_\mathrm{R} X_{S,\alpha}^y(t)
+Y_{S,\alpha\beta}^x(t)f_\beta^y(t)
\end{align}
for ${\hat{b}}=y$, where $\dot{X}_{S,\alpha}^{\hat{a}}(t)={\frac{dX_{S,\alpha}^{\hat{a}}(t)}{dt}}$ and $Y_{S,\alpha\beta}^x(t)=\int\!d\r\,r_\alpha r_\beta S_x(\r,t)$ were defined.
These equations show that the time evolution of the spin current can be expressed in terms of measurable $S_{x,y,z}(\r,t)$.

To derive Eq.~(\ref{eq_app:conductivity_SOC_result}), we evaluate spin dynamics with the linear response theory.
In general, the linear response of an operator $O(t)$ to the external force $f_\beta^{\hat{b}}(t)$ in the frequency space is given by
\begin{align}\label{eq_app:Kubo_O(t)}
\frac{\<\delta\tilde{O}(\omega)\>_{f_\beta^{\hat{b}}}}{\tilde{f}_\beta^{\hat{b}}(\omega)}&\equiv
\frac{\<\tilde{O}(\omega)\>_{f_\beta^{\hat{b}}}-\<\tilde{O}(\omega)\>_0}{\tilde{f}_\beta^{\hat{b}}(\omega)}\nonumber\\
&=i\int_0^\infty\!dt\,e^{i\omega^+ t}\<[O(t),X_{S,\beta}^{\hat{b}}(0)]\>_0,
\end{align}
where $\<\cdots\>_{f_\beta^{\hat{b}}}$ and $\<\cdots\>_0$ denote the expectation values with and without the perturbation $f_\beta^{\hat{b}}(t)$, respectively, and $\tilde{O}(\omega)=\int_{-\infty}^\infty\!dt\,e^{i\omega t}O(t)$ and $\tilde{f}_\beta^{\hat{b}}(\omega)=\int_{-\infty}^\infty\!dt\,e^{i\omega t}f_\beta^{\hat{b}}(t)$.
Substituting $O=J_{S,\alpha}$ and $\hat{b}=z$ into Eq.~(\ref{eq_app:Kubo_O(t)}), 
performing the integration by parts, and using Eqs.~(\ref{eq_app:conductivity_SOC}) and (\ref{eq_app:dot_X_S_f^z}), we obtain
\begin{align}\label{eq;J_S_SOC}
\frac{\<\delta\tilde{J}_{S,\alpha}(\omega)\>_{f_\beta^z}}{\tilde{f}_\beta^z(\omega)}
&=\sigma_{\alpha\beta}^{(S)}(\omega)-\frac{i\Omega_\mathrm{R}}{\omega^+}\nonumber\\
&\quad\times i\int_0^\infty\!dt\,e^{i\omega^+ t}\<[J_{S,\alpha}(t),X_{S,\beta}^y(0)]\>_0.
\end{align}
The last term in Eq.~(\ref{eq;J_S_SOC}) is related to Eq.~(\ref{eq_app:Kubo_O(t)}) with $O=J_{S,\alpha}$ and ${\hat{b}}=y$, leading to
\begin{align}\label{eq_app:conductivity_SOC1}
\sigma_{\alpha\beta}^{(S)}(\omega)
=\frac{\<\delta\tilde{J}_{S,\alpha}(\omega)\>_{f_\beta^z}}{\tilde{f}_\beta^z(\omega)}+\frac{i\Omega_\mathrm{R}}{\omega^+}\frac{\<\delta\tilde{J}_{S,\alpha}(\omega)\>_{f_\beta^y}}{\tilde{f}_\beta^y(\omega)}.
\end{align}
From Eqs.~(\ref{eq_app:dot_X_S_f^z}) and (\ref{eq_app:dot_X_S_f^y}), the responses of the spin current are given in terms of those of measurable quantities as
\begin{align}\label{eq_app:J_S_to_f^z}
\<\delta J_{S,\alpha}(t)\>_{f_\beta^z}
&=\<\delta\dot{X}_{S,\alpha}^z(t)\>_{f_\beta^z}-\Omega_\mathrm{R}\<\delta X_{S,\alpha}^y(t)\>_{f_\beta^z},\\
\<\delta J_{S,\alpha}(t)\>_{f_\beta^y}
&=\<\delta\dot{X}_{S,\alpha}^z(t)\>_{f_\beta^y}-\Omega_\mathrm{R}\<\delta X_{S,\alpha}^y(t)\>_{f_\beta^y}\nonumber\\
&\quad-\<Y_{S,\alpha\beta}^x\>_0f_\beta^y(t).
\end{align}
By using these equations and $\tilde{\dot{X}}_{S,\alpha}^z(\omega)=-i\omega\tilde{X}_{S,\alpha}^z(\omega)$ and defining $\Xi_{\alpha\beta}^{{\hat{a}}{\hat{b}}}(\omega)
\equiv\<\delta\tilde{X}_{S,\alpha}^{\hat{a}}(\omega)\>_{f_\beta^{\hat{b}}}/\tilde{f}_\beta^{\hat{b}}(\omega)$, Eq.~(\ref{eq_app:conductivity_SOC1}) is found to become Eq.~(\ref{eq_app:conductivity_SOC_result}).
In the case of the single-frequency driving $f_\beta^{\hat{b}}(t)=F_\beta^{\hat{b}}\cos(\omega_0t)$, we can find the expression of $\<\delta X_{S,\alpha}^{\hat{a}}(t)\>_{f_\beta^{\hat{b}}}$ as
\begin{align}
\frac{\<\delta X_{S,\alpha}^{\hat{a}}(t)\>_{f_\beta^{\hat{b}}}}{F_\beta^{\hat{b}}}
&=\frac{\Xi_{\alpha\beta}^{{\hat{a}}{\hat{b}}}(\omega_0)+\Xi_{\alpha\beta}^{{\hat{a}}{\hat{b}}}(-\omega_0)}{2}\cos(\omega_0t)\nonumber\\
&\quad+\frac{\Xi_{\alpha\beta}^{{\hat{a}}{\hat{b}}}(\omega_0)-\Xi_{\alpha\beta}^{{\hat{a}}{\hat{b}}}(-\omega_0)}{2i}\sin(\omega_0t).
\end{align}
Using $\Xi_{\alpha\beta}^{{\hat{a}}{\hat{b}}}(-\omega_0)=[\Xi_{\alpha\beta}^{{\hat{a}}{\hat{b}}}(\omega_0)]^*$ resulting from the hermiticity of $X_{S,\alpha}^{\hat{a}}$ and $X_{S,\beta}^{\hat{b}}$, we finally obtain Eq.~(\ref{eq_app:<delta_X_S>_omega_0}).

\section{\label{appendix:nonequilibrium}Generalization to nonequilibrium states}
In solid state physics, the optical charge conductivity measured in pump-probe experiments has provided valuable information on nonequilibrium phenomena such as photoinduced insulator-metal transition~\cite{Iwai:2003,Cavalleri:2004,Okamoto:2007}.
Here, we discuss the generalization of the optical spin conductivity as a probe for nonequilibrium phenomena and the method of the measurement.
As in the case of a charge response in nonequilibrium electron systems~\cite{Shimizu:2010}, we can see that the nonequilibrium optical spin conductivity $\sigma_{\alpha\beta}^{(S)}(\omega;t)$ generally depends on time $t$ but has properties similar to Eqs.~(\ref{eq:conductivity1}) and (\ref{eq:f-sum}) in equilibrium.
In addition, $\sigma_{\alpha\beta}^{(S)}(\omega;t)$ can be experimentally extracted by measuring the spin density profiles.

For simplicity, we here focus on the following experimental situation with the spin conservation:
A nonequilibrium state we are interested in is driven by time-dependent vector and scalar potentials ${\bm{A}}_{s_z}(\r;t)$ and $V_{s_z}(\r;t)$, which we call pump fields.
The pump fields can be so strong that the system can be driven far from equilibrium.
To investigate the spin current response of this nonequilibrium state, a weak spin-dependent force $f_\beta(t)$, which we call a probe field, is applied to the system in addition to the pump fields.
Specifically, the Hamiltonian in this setup consists of two terms $\H(t)=H(t)+\delta H_\beta(t)$.
The nonperturbative part $H(t)$ has the form of Eq.~(\ref{eq_app:H}) with the replacement ${\bm{A}}_{s_z}(\r)\to{\bm{A}}_{s_z}(\r;t)$ and $V_{s_z}(\r)\to V_{s_z}(\r;t)$, while $\delta H_\beta(t)$ is given by Eq.~(\ref{eq:deltaH(t)}) with $f_\beta(t)$.
Performing perturbative expansion in $f_{\beta}(t)$ in the Keldysh formalism, we can obtain the response of $\J_S(t)=\frac{d\X_S(t)}{dt}$ in the presence of the pump fields to $f_{\beta}(t)$.
The nonequilibrium optical spin conductivity is given as the corresponding response function:
\begin{subequations}\label{eq_app:neq}
\begin{align}
\<\delta J_{S,\alpha}(t)\>&=\<J_{S,\alpha}(t)\>-\<J_{S,\alpha}(t)\>_0\nonumber\\
&=\int_{-\infty}^\infty\!\!\frac{d\omega}{2\pi}\,\sigma_{\alpha\beta}^{(S)}(\omega;t)\tilde{f}_{\beta}(\omega)e^{-i\omega t},\\
\label{eq_app:conductivity0_neq}
\sigma_{\alpha\beta}^{(S)}(\omega;t)&=i\int_{-\infty}^\infty\!\!d\tau\,e^{i\omega^+ \tau}\theta(\tau)\<[J_{S,\alpha}(t),X_{S,\beta}(t-\tau)]\>_0,
\end{align}
\end{subequations}
where $\tilde{f}_{\beta}(\omega)$ is the Fourier transform of $f_{\beta}(t)$.
In this Appendix, $\<\cdots\>$ denotes the expectation values with respect to a nonequilibrium state under both pump and probe fields, while $\<\cdots\>_0$ denotes that under the pump fields in the absence of $f_\beta(t)$.
We note that, as clarified in Ref.~\cite{Shimizu:2010}, Eqs.~(\ref{eq_app:neq}) never mean the existence of the fluctuation-dissipation relation in nonequilibrium systems.
By performing integration by parts in Eq.~(\ref{eq_app:conductivity0_neq}) and using Eq.~(\ref{eq_app:[r,p]}) as well as $\J_S(t-\tau)=-\frac{d\X_S(t-\tau)}{d\tau}$ resulting from the spin conservation, we obtain
\begin{align}\label{eq_app:conductivity1_neq}
\sigma_{\alpha\beta}^{(S)}(\omega;t)=\frac{i}{\omega^+}\left(\delta_{\alpha\beta}\sum_{s_z}\frac{s_z^2N_{s_z}}{m}+\chi_{\alpha\beta}(\omega;t)\right),
\end{align}
where $\chi_{\alpha\beta}(\omega;t)=-i\int_{-\infty}^\infty\!\!d\tau\,e^{i\omega^+\tau}\,\theta(\tau)\<[J_{S,\alpha}(t),J_{S,\beta}(t-\tau)]\>_0$.
The causality condition in $\chi_{\alpha\beta}(\omega;t)$ leads to the Kramers-Kronig relation as in Eq.~(\ref{eq_app:Kramers-Kronig})~\cite{Shimizu:2010}.
As a result, we obtain the following $f$-sum rule~\footnote{While Ref.~\cite{Shimizu:2010} considers the response of a current density, whose $f$ sum generally depends on time, we now consider the response of the total spin current and thus the $f$ sum is related to $N_{s_z}$ independent of time.}:
\begin{align}\label{eq_app:f-sum_neq}
\int_{-\infty}^{\infty}\!\!\frac{d\omega}{\pi}\Re\,\sigma_{\alpha\beta}^{(S)}(\omega;t)=\delta_{\alpha\beta}\sum_{s_z}\frac{s_z^2 N_{s_z}}{m}.
\end{align}
Equations~(\ref{eq_app:conductivity1_neq}) and (\ref{eq_app:f-sum_neq}) are similar to Eqs.~(\ref{eq:conductivity1}) and (\ref{eq:f-sum}) except for the time dependence of $\sigma_{\alpha\beta}^{(S)}(\omega;t)$.

The optical spin conductivity for the nonequilibrium state can be experimentally extracted by applying a single-frequency probe field $f_\beta(t)=F_\beta\cos(\omega_0t+\phi)$ with a phase $\phi$~\cite{Shimizu:2010}.
Because of the time dependence of $\sigma_{\alpha\beta}^{(S)}(\omega;t)$, a simple relation of $\sigma_{\alpha\beta}^{(S)}(\omega_0;t)$ to $\<\delta X_{S,\alpha}(t)\>$  [Eq.~(\ref{eq:X_S(t)})] is lost unless the state driven by the pump fields can be regarded as a nonequilibrium steady state.
On the other hand, a relation to the spin current similar to Eq.~(\ref{eq_app:oscillating_J_S(t)}) generally holds:
\begin{align}\label{eq_app:oscillating_J_S(t)_neq}
\frac{\<\delta J_{S,\alpha}(t)\>}{F_\beta}&=\Re\,\sigma_{\alpha\beta}^{(S)}(\omega_0;t)\cos(\omega_0t+\phi)\nonumber\\
&\quad+\Im\,\sigma_{\alpha\beta}^{(S)}(\omega_0;t)\sin(\omega_0t+\phi).
\end{align}
From this relation, $\sigma_{\alpha\beta}^{(S)}(\omega_0;t)$ can be rewritten as~\cite{Shimizu:2010}
\begin{align}
\sigma_{\alpha\beta}^{(S)}(\omega_0;t)=\frac{\<\delta J_{S,\alpha}(t)\>|_{\phi=0}-i\<\delta J_{S,\alpha}(t)\>|_{\phi=\pi/2}}{F_\beta e^{-i\omega_0t}}.
\end{align}
Because of $\<\delta J_{S,\alpha}(t)\>=\frac{d\<\delta X_{S,\alpha}(t)\>}{dt}$, the nonequilibrium optical spin conductivity can be experimentally determined by measuring two kinds of responses $\<\delta X_{S,\alpha}(t)\>|_{\phi=0,\pi/2}$ with fixed pump fields.


\begin{thebibliography}{99}
\bibitem{Note:Linear_response}
For systems with inversion invariance including superfluids examined later, the second-order term exactly vanishes and thus the correction to Eq.~(\ref{eq:conductivity}) is of third order in $f_\beta(t)$.
\bibitem{Note:S=1/2}
Note that our definition of the optical spin conductivity with $S=1/2$ is different from that in Refs.~\cite{Enss:2012,Enss:2013} by a factor $1/4$.
\bibitem{Note:mu}
Within the mean-field theory at zero temperature, $\mu$ changes its sign at $(k_\mathrm{F}a)^{-1}=4 \Gamma^2\left(\frac{3}{4}\right)/[\pi^2\Gamma\left(\frac{1}{4}\right)]^{2/3}=0.5531$, where $\Gamma(z)$ is the gamma function.
\bibitem{Note:mean-field}
For example, at zero temperature and unitarity, the contact $C/k_{\rm F}^4=0.118$ within the mean-field theory~\cite{Ohashi} is close to the recent experimental results $C_{\rm exp.}/k_{\rm F}^4\simeq 0.1$~\cite{Horikoshi,Carcy,Mukherjee}.
\bibitem{Note:Bogoliubov}
This theory is available in a weakly-interacting BEC near zero temperature~\cite{kawaguchi}.
\bibitem{Note:TLL}
We implicitly assume that the Umklapp process yielding a term proportional to $\cos(\sqrt{8}\phi_C)$ is absent, provided that the density is away from half filing. However, it is irrelevant to this work.
\end{thebibliography}
\end{document}